\documentclass[%
 reprint,
 superscriptaddress,
nofootinbib,
 amsmath,amssymb,
 aps,
floatfix,
]{revtex4-2}
\usepackage{float} 
\usepackage{subcaption}
\usepackage{siunitx}
\usepackage{graphicx}% Include figure files
\usepackage{dcolumn}% Align table columns on decimal point
\usepackage{bm}% bold math
\newcommand{\gm}{$g$-mode }
\newcommand{\bv}{Brunt-V\"ais\"al\"a }

\begin{document}

\preprint{APS/123-QED}

%\title{G-Mode Oscillations in Relativistic Mean Field Models in Hyperonic Matter}% Force line breaks with \\

\title{$g$-mode Oscillations in Neutron Stars with Hyperons} 

\author{Vinh Tran}
\email[Corresponding author:~]{vinh.tran02@student.csulb.edu}
\affiliation{Department of Physics and Astronomy, California State University Long Beach, Long Beach, California 90840, USA}
\author{Suprovo Ghosh}
\email{suprovoh@iucaa.in}
\affiliation{Inter-University Centre for Astronomy and Astrophysics,
Pune University Campus,
Pune 411007, India}
\author{Nicholas Lozano}
\email{Nicholas.Lozano@csulb.edu}
\affiliation{Department of Physics and Astronomy, California State University Long Beach, Long Beach, California 90840, USA}
\author{Debarati Chatterjee}
\email{debarati@iucaa.in}
\affiliation{Inter-University Centre for Astronomy and Astrophysics,
Pune University Campus,
Pune 411007, India}
\author{Prashanth Jaikumar}
\email{prashanth.jaikumar@csulb.edu}
\affiliation{Department of Physics and Astronomy, California State University Long Beach, Long Beach, California 90840, USA}

\date{\today}

%%%%%%%%%%%%%%%%%%%%%%%%%%%%%%%%%%%%%%%%%%%%%%%%%%
\begin{abstract}
%%%%%%%%%%%%%%%%%%%%%%%%%%%%%%%%%%%%%%%%%%%%%%%%%%
A common alternative to the standard assumption of nucleonic composition of matter in the interior of a neutron star is to include strange baryons, particularly hyperons. Any change in composition of the neutron star core has an effect on $g$-mode oscillations of neutron stars, through the compositional dependence of the equilibrium and adiabatic sound speeds. We study the core $g$-modes of a neutron star contaning hyperons, using a variety of relativistic mean field models of dense matter that satisfy observational constraints on global properties of neutron stars. Our selected models predict a sharp rise in the $g$-mode frequencies upon the onset of strange baryons. Should $g$-modes be observed in the near future, their frequency could be used to test the presence of hyperonic matter in the core of neutron stars.
%%%%%%%%%%%%%%%%%%%%%%%%%%%%%%%%%%%%%%%%%%%%%%%%%%
\end{abstract}
%%%%%%%%%%%%%%%%%%%%%%%%%%%%%%%%%%%%%%%%%%%%%%%%%%

\maketitle

%\tableofcontents 

%%%%%%%%%%%%%%%%%%%%%%%%%%%%%%%%%%%%%%%%%%%%%%%%%%

\author{Vinh Tran}
 \affiliation{}%Lines break automatically or can be forced with \\
%  \email{Second.Author@institution.edu}
% \affiliation{%
%  Authors' institution and/or address\\
%  This line break forced with \textbackslash\textbackslash
% }%

\collaboration{}%\noaffiliation

%  \homepage{http://www.Second.institution.edu/~Charlie.Author}
% \affiliation{
%  Second institution and/or address\\
%  This line break forced% with \\
% }%
% \affiliation{
%  Third institution, the second for Charlie Author
% \affiliation{%
%  Authors' institution and/or address\\
%  This line break forced with \textbackslash\textbackslash
% }%

\collaboration{}%\noaffiliation

\date{\today}% It is always \today, today,

% \begin{abstract}
% G-mode oscillations using relativistic mean field models for hyperonic and hybrid quark star matter (maybe) using different sets of models generalizing a method for calculating the sound speed difference 
% \end{abstract}

\maketitle

%\tableofcontents

\section{Introduction}

% \begin{itemize}
%     \item Talk about current state of affairs. Mention really large neutron stars? Mention GW 190814 secondary objefct?
    
%     \item Gravitational Waves and LIGO 
    
%     \item Potential for addressing a new problem of core composition and constraints using G-mode oscillations
    
%     \item Unsolved problems regarding nuclear theory at neutron star densities, QCD phase diagram, and neutron stars in general 
% \end{itemize}

% The objective of this paper is to demonstrate a method for calculating principal $g$-mode oscillations based upon finding the sound speed difference $c_s^2 - c_e^2$ in a manner that easily generalizes between different models in the overall family of relativistic mean field models. In doing so, we calculate and investigate the $g$-mode oscillations for a variety of different models as well as compositions to explore the dependence of the $g$-mode oscillation frequency spectrum on composition to compare and contrast nucleonic and hyperonic matter as well as nucleonic-hyperonic and quark transitions (maybe?).

The composition of matter in the interior of a neutron star, uncertain at present, is relevant to fundamental questions about the phase of strongly interacting, cold and dense matter~\cite{Zhao_2020,2021arXiv211212157D,2020JPhCS1602a2013D}. An equation of state, which relates state variables in thermodynamic equilibrium, may be derived from a theoretical model of purely nucleonic matter (npe or npe$\mu$) \cite{Huang_2020,Thapa_2021,Clevinger_Dexheimer}, hyperonic matter (npe$\mu$Y)~\cite{glendenning_1985,Glendenning_1991,Tu_2022,Thapa_2021, Dexheimer_2021,Clevinger_Dexheimer}, matter with Bose condensates or delta baryons~ \cite{Dexheimer_2021_delta,Thapa_2021}, or hybrid matter with a phase transition from nucleonic to quark degrees of freedom~\cite{Zhao_2020,Dexheimer_2021,Clevinger_Dexheimer} to name a few possibilities. One way to test various theoretical models of dense matter is to compare predicted macroscopic properties of neutron stars with astronomical observations. For example, the appearance of hyperons can alter a neutron star's maximum mass, radius, cooling or gravitational wave (GW) emission from unstable quasi-normal modes compared to the purely nucleonic scenario~\cite{VidanaEPJA}.  

Theoretical models of neutron stars must satisfy maximum mass constraints gleaned from, for eg., observations of the ``black widow" pulsar PSR J0952-0607, the heaviest neutron star to date with mass of $2.35_{-0.17}^{+0.17}\,M_\odot$\cite{Romani_2022} or the suggested secondary component in the binary merger event GW190814~\cite{Abbott_2020} with mass of 2.5$M_{\odot}$ or higher. The so-called ``hyperon puzzle" refers to the softening effect of hyperons that makes such constraints hard or impossible to satisfy~\cite{2017hspp.confj1002B}, though many solutions have been proposed~\cite{Kolomeitsev_2016,2021PhRvC.103c5810P,2022ApJ...925...16T}. Recent observational constraints from the Neutron star Interior Composition Explorer collaboration (NICER) report a mass of $1.34_{-0.16}^{+0.15}\,M_\odot$ and radius of $12.71^{+1.14}_{-1.19}\,\text{km}$ from~\cite{Riley_2019} and~\cite{Miller_2019} report $1.44^{+0.15}_{-0.14}\,M_\odot$ and radius of $13.02_{-1.06}^{+1.24}\,\text{km}$ for the same star. Similarly, NICER observations of PSR J0740+6620 yield a mass of $2.08_{-0.07}^{+0.08}\,M_\odot$ with equatorial radius of $13.7_{-1.5}^{+2.6}\,\text{km}$ from \cite{Miller_2021} and $2.072_{-0.066}^{+0.067}\,M_\odot$ with radius $12.39_{-0.98}^{+1.30}\,\text{km}$ from \cite{Riley_2021}. Gravitational wave observations from compact binary merger events such as GW170817~\cite{Abbott_2017a} and GW190814~\cite{Abbott_2020} are another probe of the equation of state~\cite{2019PrPNP.10903714B}. Assuming that the secondary object in GW190814 is a heavy neutron star, the analysis in~\cite{Abbott_2020} yields the tidal deformability $\Lambda_{1.4}$ of a canonical mass neutron star to be ${616}_{-158}^{+273}$. 

While these constraints are narrowing the allowed range of neutron star mass and radius, it is difficult to draw firm conclusions on, or distinguish between, different interior compositions based on static global properties of neutron stars alone~\cite{Wei_2019,Wei2018-ag,Alford_2005}. Although the presence of non-nucleonic species such as hyperons or phase transitions to quark matter tend to lead to a softening of the equation of state and some tension with astrophysical constraints~\cite{2022FrASS...964294G,Vidana_2018, Chatterjee_2016, Bedaque_2015}, there are still many models that satisfy current astrophysical constraints~\cite{2022arXiv220905699R,Constantinou_2021,2017ApJ...834....3T}. A different approach, namely that of stellar oscillations, may provide a new tool for addressing the problem of composition more directly. The secular quasi-normal oscillation modes of neutron star carry information about the interior composition and viscous forces that damp these modes~\cite{Cowling,Schmidt,Thorne}. Examples include the fundamental $f$-mode, $p$-modes and $g$-modes (driven by pressure and buoyancy respectively), as well as $r$-modes (Coriolis force) and pure space-time $w$-modes. Several of these modes may be excited during a supernova explosion, or in isolated perturbed neutron stars or during the post-merger phase of a binary NS ~\cite{Kokkotas2001,Stergioulas2011,Vretinaris2020}. Spin and eccentricity may enhance the excitation of the $f$-modes during the inspiral phase of a neutron star merger ~\cite{Chirenti_2017,Steinhoff2021}. The fundamental $f$-modes as well as composition-driven $g$-modes are within the sensitivity range of current generation of GW detectors and the former is correlated with the tidal deformability ~\cite{Chan2014,Hinderer2016,Pratten2020,PhysRevD.101.103009}. 
\\

Our focus in this work will be on $g$-modes of hyperonic stars. It is known that $g$-modes are particularly sensitive to composition, as shown in studies ranging from $npe$ and $npe\mu$ matter~\cite{Wei2018-ag} to hybrid stars exhibiting a first order phase transition from nucleonic matter to a deconfined quark phase~\cite{PJ-PRD} or in a crossover model \cite{Constantinou_2021}.  It was found that the appearance of quarks in neutron star matter, especially via a first order transition, leads to a dramatic increase in the $g$-mode oscillation frequency. 
In this work, we extend this analysis to consider compositions including hyperons as well ($npe\mu$Y).
\\

This paper is organized as follow: In sec. \ref{background} we introduce the theoretical framework for $g$-mode oscillation followed by a discussion of two sound speeds $c_s^2$ and $c_e^2$, whose difference drives the $g$-mode, in sec.\ref{sound_speed_diff}. In sec. \ref{rmf}, we introduce the relativistic mean field models (RMF) we sample for our calculations, phenomenological models that treat baryons as fundamental fields interacting via mesons \cite{walecka_1974, glendennning_compact_stars, glendenning_1985}. In sec. \ref{sec:adiabatic_sound_speed}, we outline the method we use for calculating the adiabatic sound speed via the sound speed difference expression introduced in sec. \ref{sound_speed_diff}. In sec. \ref{sec:results} we present our results for the $g$-mode oscillation frequencies, followed by our conclusions in sec. \ref{sec:conclusions} and an instructive derivation on sound speeds in the Appendix \ref{sound_speed_diff_proof}.

\section{$g$-Mode Oscillations}\label{background} 

In the general theory of linearized non-radial oscillations of an ideal self-gravitating fluid comprising a compact star,  the oscillatory fluid displacement of a mode with quantum numbers ${nlm}$ is represented by a vector field ${\vec{\xi}}^{nlm}(\vec{r},t)$, conveniently separable in a spherically symmetric background into radial and tangential components $\xi_r^{nlm}(\vec{r},t)$ = $\eta_r^{nl}(r)Y_{lm}(\theta,\phi){\rm e}^{-i\omega t}$ and $\xi_{\perp}^{nlm}(\vec{r},t)$ = $r\eta_{\perp}^{nl}(r)\nabla_{\perp}Y_{lm}(\theta,\phi){\rm e}^{-i\omega t}$ respectively where $Y_{lm}(\theta,\phi)$ are the spherical harmonics.     From the perturbed (Newtonian) continuity equation for the fluid, the corresponding pressure perturbation is $\delta p/\rho$ = $\omega^2r\eta_{\perp}(r)Y_{lm}(\theta,\phi){\rm e}^{-i\omega t}$, where $\rho$ is the energy density. The equations of motion (Euler equation) to be solved to determine the frequency $\omega_{nl}$ of a particular $nl$ mode (degenerate in $m$ for non-rotating stars) is
\begin{eqnarray}
\label{eq:1}
    \frac{\partial}{\partial r}(r^2\xi_r) &=&\left[\frac{l(l+1)}{\omega^2}-\frac{r^2}{c_s^2}\right]\left(\frac{\delta p}{\rho}\right) \,\,,\\
\frac{\partial}{\partial r}\left(\frac{\delta p}{\rho}\right) &=& \frac{\omega^2-N^2}{r^2}(r^2\xi_r)+\frac{N^2}{g}\left(\frac{\delta p}{\rho}\right) \,\,.
\label{eq:2}
\end{eqnarray}
 where we have suppressed the indices on $\omega$ and $\xi$ and $N^2= \frac{c_s^2 - c_e^2}{c_s^2 c_e^2}$ is the \bv\, frequency. For a given equation of state (stellar structure), a global solution of the linear perturbation equations, eqns. (\ref{eq:1}) and (\ref{eq:2}), is found subject to  boundary conditions of regularity at the stellar center ($r$ $\rightarrow$ 0) and vanishing of the Lagrangian pressure variation~\footnote{The Lagrangian variation of a fluid variable is related to the Eulerian variation through the operator relation $\Delta \equiv \delta + \xi\cdot\nabla$.} $\Delta p$ = $c_s^2 \Delta \rho$ at the surface. These solution values represent the discrete $g$-mode spectrum for a chosen stellar model. 
As in other works~\cite{Constantinou_2021, PJ-PRD, Kantor_2014, Reisenegger_1992, McDermott_1983}, we use the Cowling approximation~\cite{1942Obs....64..224C} which neglects the back reaction of the gravitational potential, while extending eqns. (\ref{eq:1} and (\ref{eq:2}) to include the relativistic effects of the matter~\cite{Kantor_2014} which yields
 \begin{align}
    - \frac{1}{e^{\lambda/2}r^2}\frac{\partial}{\partial r}[e^{\lambda/2} r^2 \xi_r] + \frac{\ell(\ell+1) e^\nu}{r^2\omega^2}\frac{\delta p}{p +\varepsilon} - \frac{\Delta p}{\gamma p} &= 0\\
    \frac{\partial \delta p}{\partial r} + g\left(1 + \frac{1}{c_s^2}\right)\delta p + e^{\lambda - \nu}h(N^2 - \omega^2)\xi_r &= 0
\end{align}
where $N^2$, the \bv\, frequency is slightly modified to
\begin{align}
\label{bvf}
    N^2 &= g^2 \left(\frac{c_s^2 - c_e^2}{c_s^2 c_e^2}\right)e^{\lambda - \nu} \,.
\end{align}
where $\nu(r)$ and $\lambda(r)$ are metric functions of the unperturbed star which feature in the Schwarzschild {\it interior} metric, and $\gamma=(n_B/p)\partial p(n_B,Y_p)/\partial n_B$ is the adiabatic index with $n_B$ the baryon density.

The impact of the Cowling approximation, compared to a full general relativistic calculation typically only affects the frequencies of the \gm\, at the 5-10\% level~\citep{Grig} and for heavier stars~\cite{Zhao:2022toc}, therefore it does not change our conclusions qualitatively. Because we have employed the Cowling approximation and ignored the perturbations of the metric that must accompany fluid perturbations, we cannot compute the imaginary part of the eigenfrequency (damping time) of the $g$-mode~\footnote{The damping time of $g$-modes due to viscosity and gravitational wave emission, estimated  in some works~\cite{1999MNRAS.307.1001L, Wei2018-ag}, suggests that the $g$-mode can become secularly unstable for temperatures $10^8~{\rm K}<T<10^9~{\rm K}$ for rotational speeds exceeding twice the $g$-modefrequency of a static star.}. 
 
Equations (\ref{eq:1}) and (\ref{eq:2}) can be analyzed in the short-wavelength limit ($kr\gg 1$) where the local dispersion relation has two distinct branches, with the lower frequency branch corresponding to the $g$-modes. The local $g$-mode frequency is then $\omega^2\propto {\rm e}^{\lambda}N^2$~\cite{1983ApJ...268..837M}, highlighting the importance of the two sound speeds (in particular, the difference of their inverse squares, as in Eq.(\ref{bvf})). The global g-mode frequency is constant for a given stellar configuration (fixed gravitational mass $M$ and radius $R$) and can be thought of as an average of the local g-modes (although it is still sensitive to phase transitions).

In this work, we study the fundamental $g$-mode with $n$ = 1 and fix the mode's multipolarity at $l$ = 2. This is because the $l$ = 2 mode is quadrupolar in nature, and can couple to gravitational waves. Higher $l$ values (octupole and higher) are generally weaker than the quadrupole. 
The reason to study the $n$ = 1 (fundamental) $g$-mode is that the local dispersion relation for $g$-modes $\omega^2\propto 1/k^2$ implies that the $n$ = 1 excitation has the highest frequency, whereas higher values of $n$ are known to have a smaller amplitude of excitation and a weaker tidal coupling coefficient~\cite{2021PhRvD.104l3032C}. The fundamental $g$-mode is also within the sensitivity range of current generation of gravitational wave (GW) detectors{{~\cite{1994MNRAS.270..611L,2022arXiv220403037Z}}}.

% This can be re-written as
% \begin{align}
%      \frac{\partial}{\partial r}[r^2 e^{\lambda/2}\xi_r] &=
%         \frac{e^{\lambda/2}}{p+\varepsilon}\left(\frac{\ell(\ell+1)e^\nu}{\omega^2} - \frac{r^2}{c_s^2}\right)\delta p \\
%         &\qquad -  \frac{dp}{dr}\frac{1}{ c_s^2 (p+\varepsilon)} e^{\lambda/2}r^2 \xi_r\\
%         \frac{\partial \delta p}{\partial r} &= - g \left(1 + \frac{1}{c_s^2}\right)\delta p\\
%         &\qquad - 
%         \frac{e^{\lambda/2 - \nu}}{r^2}(p + \varepsilon) (N^2 - \omega^2) (e^{\lambda/2}r^2 \xi_r) 
% \end{align}

\section{Sound Speed Difference} \label{sound_speed_diff}
A necessary quantity for calculating $g$-mode oscillations is the \bv\, frequency which is proportional to the difference of the squares of two sound speeds: $c_s^2 - c_e^2$ where the equilibrium sound speed $c_e^2$ is the total derivative of the pressure $p$ with respect to energy density $\varepsilon$ and the adiabatic sound speed $c_s^2$ is the partial derivative of $p$ with respect to $\varepsilon$ while holding the composition of the matter $\mathcal{\chi}$ fixed, 
\begin{align}
    c_e^2 := \frac{dp}{d\varepsilon} \qquad \qquad c_s^2 := \frac{\partial p}{\partial \varepsilon}\bigg|_\chi 
\end{align}
where $\chi$ is shorthand denoting various particle fractions $x_i := n_i/n_B$ fixed. For baryonic compositions ($npe$ or $npe\mu$), this means fixing the proton and electron/muon fraction fixed, and is more involved for compositions with hyperons: $\Lambda^0, \Sigma^-, \Sigma^0,\Sigma^+, \Xi^-,\Xi^0$. The expressions for $p$ and $\varepsilon$ is model dependent, but encapsulates contributions from all particles present. 

Starting from the definitions of $c_s^2$ and $c_e^2$, the sound speed difference $c_s^2 - c_e^2$ can be written in terms of partial derivatives of specific linear combinations of chemical potentials $\tilde{\mu}_i$ defined in eqns. (\ref{mu_tilde}-\ref{mu_tilde_2}) as shown for $npe$ and $npe\mu$ matter in \cite{PJ-PRD}. A natural generalization of that expression to arbitrary compositions in eqn. \ref{cs2_ce2} is (as shown in Appendix A) \ref{sound_speed_diff_proof}
\begin{align}
    c_s^2 - c_e^2  = 
    \frac{n_B^2}{\mu_n} \sum_i 
    \frac{\partial \tilde{\mu}_i}{\partial n_B}\bigg|_\chi \frac{dx_i}{dn_B} \label{cs2_ce2}
\end{align}
where $\mu_n$ is the neutron chemical potential, $x_i$ is the particle fraction for the $i$th independent particle, and $\tilde{\mu}_i$ is a linear combination of chemical potentials satisfying $\tilde{\mu_i} = 0$ in $\beta$-equilibrium as defined in eqns. (\ref{mu_tilde}) and (\ref{mu_tilde_2}) below. The sum over $i$ accounts for each individual $\beta$-equilibrium condition in our system where in our case $i \in p, \Lambda^0, \Sigma^-, \Sigma^0, \Sigma^+, \Xi^-,\Xi^0$. In essence, Eq.(\ref{cs2_ce2}) provides a method for calculating the adiabatic sound speed $c_s^2$ from finding the equilibrium sound speed $c_e^2$ and sound speed difference $c_s^2 - c_e^2$ separately. 

At zero temperature and deleptonized matter, we have
\begin{align}
    \tilde{\mu}_i = \mu_n - q_i \mu_e - \mu_i \qquad &i \in \text{Baryon} \label{mu_tilde}\\
    \tilde{\mu}_\ell = \mu_e - \mu_\ell \qquad &\ell \in \text{Lepton} \label{mu_tilde_2}
\end{align}
where $q_i$ is the charge of the baryon (in units where $q_e = e = 1$) \cite{glendenning_1985} and $\mu_e$ is the electron chemical potential. For the baryon octet, the various $\tilde{\mu}_i$ are explicitly given by
\begin{align}
    \tilde{\mu}_p &= \mu_n - \mu_e - \mu_p &\tilde{\mu}_{\Lambda^0} = \mu_n - \mu_{\Lambda^0}\\
    \tilde{\mu}_{\Sigma^0} &= \mu_n - \mu_{\Sigma^0} &\tilde{\mu}_{\Xi^0} = \mu_n - \mu_{\Xi^0}\\
    \tilde{\mu}_{\Sigma^-} &= \mu_n + \mu_e - \mu_{\Sigma^-} &\tilde{\mu}_{\Xi^-} = \mu_n + \mu_e - \mu_{\Xi^-}\\
    \tilde{\mu}_{\Sigma^+} &= \mu_n - \mu_e - \mu_{\Sigma^+}
\end{align}
and for the muon $\tilde{\mu}_\mu$ = $\mu_e - \mu_\mu$\,.
%In $\beta$ equilibirum at zero temperature, all $\tilde{\mu}$=0~\cite{Rather_2021, schmitt} so
%\begin{align}
  %  \mu_n &= \mu_{\Sigma^0} = \mu_{\Xi^0}\\
  %  \mu_p &= \mu_{\Sigma^+} = \mu_n - \mu_e\\ 
  %  \mu_{\Sigma^-} &= \mu_{\Xi^-} = \mu_n + \mu_e 
%\end{align}
% In this context, it is natural to then take the neutron and electron to be the dependent particles with all other baryons and leptons to be the independent particles. This has a secondary but equally important effect of reducing the degrees of freedom making the system of equations that appear in Sec. \ref{rmf} consistent.
Physically, the sound speed difference is a quantitative measure of the restoration of chemical equilibrium when a perturbation occurs. As the $g$-mode frequency is dependent on $c_s^2 - c_e^2$, it follows that if $dx_i/dn_B$ is large, i.e., when new species enter the system, the $g$-mode frequency will change sharply. Indeed, that is what we find in our models, as elaborated below.

\section{Model for neutron star structure \label{rmf}}
To model the matter in the core of the star, we use relativistic mean field models (RMF) ~\cite{Han_2019, Oertel_2017, glendenning_1985, walecka_1974} which are particularly well suited for calculating the adiabatic sound speed via the method described in Sec. \ref{sound_speed_diff}. 
%We find a general method for calculating the adiabatic sound speed $c_s^2 = \partial P/\partial \epsilon|_\chi$, a quantity needed to calculate the $g$-mode oscillation frequency in the framework for a general RMF models. 
Specifically, we sample six different RMF models with a variety of different baryon-meson and meson-meson interactions. Four of these are nonlinear relativistic mean field models (NLRMF): GM1-Y5 \cite{Glendenning_1991, Oertel_2015}, Big Apple \cite{Fattoyev_2020, Das_2021}, and Hornick 65, 70 \cite{Hornick_2018}. The remaining two are density dependent relativistic mean field models (DDRMF): DD-MEX \cite{Taninah_2020, Tu_2022, Thapa_2021, Huang_2020}, and DD-ME2 \cite{lalazissis_2005, Tu_2022, Thapa_2021, Huang_2020}. For the most part, these models were originally formulated and provided to model $npe\mu$ matter. We extend these models to include hyperons via a standard SU(6) symmetry argument and fits to hyperonic optical potentials to generate the meson coupling constants.  \cite{Pradhan_2022, Miyatsu_2013, Thapa_2021, Oertel_2015}. 
The equations of motion from the model's Lagrangian, subject to local (charge neutrality) and global conservation laws (baryon number conservation) can be solved for any desired baryon/meson field as a function of baryon density $n_B$ and compositions $\chi$, which then allows for calculating chemical potential derivatives. $\beta$-equilibrium is then imposed to determine all particle fractions as a function of $n_B$ only. 

\subsection{Nonlinear Relativistic Mean Field Models \label{sec:NLRMF}}
The first class of models that we consider are nonlinear relativistic mean field models (NLRMF) that describe baryon-meson  interactions with various mesons such as the isoscalar-scalar $\sigma$, isoscalar-vector $\omega$, isovector-vector $\rho$. Models that include hyperons can also include other strange meson degrees of freedom - the hidden strangeness isoscalar-vector $\phi$, the isovector-vector $\delta$ and or the isoscalar-scalar $\xi$ mesons \cite{Tu_2022}. Specifically, the NLRMF models we use in this work are Big Apple \cite{Fattoyev_2020}, and Hornick 65 and 70 models \cite{Hornick_2018}, with the various baryon-meson and meson-meson coupling constants listed in Table \ref{tab:nonlinear_rmf_models}. These models chosen differ from one another primarily in their baryon-meson and meson-meson interactions. However, as shown in Fig. (\ref{fig:mass_radius}) and Fig. (\ref{fig:tidal_deformability_linear}) neutron stars described by these models satisfy current. As a result, we can investigate possible $g-$mode dependence on interaction specific terms.

The specific form of meson-meson interactions may vary from one model to another, but the most general Lagrangian can be split into the following terms 
\begin{align}
    \mathcal{L} &= \mathcal{L}_\text{B}^{\text{kin}} + \mathcal{L}_\ell^{\text{kin}} + \mathcal{L}_\text{M}^{\text{kin}} + \mathcal{L}_\text{int} - U_\text{NL} 
\end{align}
where the kinetic mesonic Lagrangian explicitly is
\begin{equation} 
\begin{aligned}
    \mathcal{L}^{\text{kin}}_\text{M} &= 
    \frac{1}{2}(\partial^\mu \sigma \partial_\mu \sigma - m_\sigma^2 \sigma^2) + 
    \frac{1}{2}(\partial^\mu \bm{\delta} \partial_\mu \bm{\delta} - m_\delta^2 \bm{\delta}^2)\\
    &\quad - \frac{1}{4}W^{\mu\nu}W_{\mu\nu} + 
    \frac{1}{2}m_\omega^2 \omega^\mu \omega_\mu  - 
    \frac{1}{4}\bm{R}^{\mu\nu}\bm{R}_{\mu\nu}\\
    &\quad + 
    \frac{1}{2}m_\rho^2 \bm{\rho}^\mu \bm{\rho}_\mu - 
    \frac{1}{4}\Phi^{\mu\nu}\Phi_{\mu\nu} + 
    \frac{1}{2}m_\phi^2 \phi^\mu \phi_\mu\\
    &\quad + \frac{1}{2}(\partial_\mu \xi \partial^\mu \xi - m_\xi^2 \xi^2)
\end{aligned}
\end{equation}
with $W^{\mu\nu} = \partial^\mu \omega^\nu - \partial^\nu \omega^\mu$ and $\bm{R}^{\mu\nu} = \partial^\mu \bm{\rho}^\nu - \partial^\nu \bm{\rho}^\mu$, $\Phi^{\mu\nu}  = \partial^\mu \phi^\nu - \partial^\nu \phi^\mu$. In the mean field approximation, where spatial variations of the meson fields are neglected and the meson fields are replaced by their ground state expectation value \cite{glendenning_1985}, $\mathcal{L}_\text{int}$, which describes the baryon-meson interaction, takes the form 
\begin{equation} 
\begin{aligned}
    \mathcal{L}_\text{int} &= - \sum_i \bar{\psi}_i [\gamma_0(g_{\omega i}\omega + g_{\rho i}I_{3i}\rho + g_{\phi i}\phi)\\
    &\qquad - (m_i - g_{\sigma i}\sigma - g_{\delta i}I_{3i}\delta - g_{\xi i}\xi)]\psi_i 
\end{aligned}
\end{equation} 
where $\psi_i$ and $m_i$ are the $i$th baryon field and bare mass respectively, the $g_{\alpha i}$ for $\alpha \in \omega,\rho,\phi,\xi$ are the coupling constants coupling baryons to mesons, $I_{3i}$ gives the isospin projection of the $i$th baryon species, and $\omega,\rho,\phi,\sigma,\delta,\xi$ represent the mean field expectation value of the meson fields. Similarly, $U_{NL}$, which describes the meson-meson interactions, is given by 
\begin{equation} 
\begin{aligned}
    U_{NL} &= 
    \frac{1}{3}b m_N (g_{\sigma N}\sigma)^3 + 
    \frac{1}{4}c (g_{\sigma N}\sigma)^4\\
    &\qquad - \Lambda_\omega g_{\rho N}^2 g_{\omega N}^2 \omega^2 \rho^2 - \frac{\xi_\omega}{4!}g_{\omega N}^4 \omega^4  \label{eqn:U_NL}
\end{aligned}
\end{equation} 
where $m_N$ is the nucleon bare mass, $b,c,\Lambda_\omega,\xi_\omega$ are coupling constants. There are three main interactions of note: a cubic and quartic self interaction of the $\sigma$ mesons~\cite{glendenning_1985, schmitt},  a quartic $\omega^2\rho^2$ interaction between $\omega$ and $\rho$ mesons and a quartic $\omega^4$ self interaction. For an arbitrary Lagrangian of this form, we can identify the chemical potential of a baryon $\mu_i$ from the interaction Lagrangian \cite{Hornick_2018, glendenning_1985, schmitt}
\begin{align}
    \mu_i^* &= E_{F_i}^* = \mu_i - g_{\omega i}\omega - g_{\rho i}I_{3i}\rho - g_{\phi i}\phi 
\end{align}
where
\begin{align}
    E_{F_i}^* = \sqrt{k_{F_i}^2 + {m_i^*}^2} \label{eqn:effective_energy}
\end{align}
and where $k_{F_i}$ is the Fermi momenta related to the various fermionic number densities $n_i$ (i.e., the vector number density $n_i$) by 
\begin{align}
    n_i := \langle \psi_i^\dagger \psi_i\rangle = \int_0^{k_{F_i}}  \frac{d^3k}{(3\pi^2)} = \frac{k_{F_i}^3}{3\pi^2}
\end{align}
and the scalar density $n_i^s$ is given by \cite{glendenning_1985}
\begin{align}
    n_i^s := \langle \bar{\psi}_i\psi_i \rangle 
    &= \frac{1}{\pi^2}
    \int_0^{k_{F_i}} \frac{m_i^*}{E_{F_i}^*}k^2\,dk\\
    &= \frac{m_i^*}{2\pi^2} 
    \left[k_{F_i}E_{F_i}^* - {m_i^*}^2\ln \frac{k_{F_i} +E_{F_i}^*}{m_i^*}\right] \label{ns1}
\end{align}

\begin{table}
    \begin{centering} 
    \begin{tabular}{ccccc}
    \hline 
    \hline 
      Model   & GM1-Y5 &Hornick 65 &Hornick 70 &Big-Apple  \\
    \hline 
     \hline 
      $m_\sigma$ (MeV) &550.0  &550.0 &550.0 &492.730\\ 
      \hline 
        $n_0$ (fm$^{-3}$) & 0.153  &0.150 & 0.150 &0.155\\ 
        $b$ &0.002947 &-0.00198839 &-0.004315 &0.005280\\
        $c$ &-0.001070 &-0.0028455 &-0.004347 &-0.003623\\ 
        $\Lambda_\omega$ &0.0  &0.0295148 &0.031432 &0.047471\\
        $\xi_\omega$ &0.0 &0.0 &0.0 &0.00070\\
    \hline 
       $g_{\sigma N}$  &  9.57 &10.4291 &9.84608 &9.6699\\
       $g_{\omega N}$ & 10.61 &11.7742 &10.7467 &12.3116   \\ 
       $g_{\rho N}$ & 8.20 &10.1865 & 9.9829 &14.1618\\ 
    \hline 
    $g_{\sigma \Lambda}$ &5.84 &5.898 &5.99472 &5.7656\\
    $g_{\sigma \Sigma}$ &3.87 &3.991 &3.8988775 &4.1314\\ 
    $g_{\sigma \Xi}$ &3.06 &2.949 &3.10215 &2.8556 \\ 
    \hline 
    $g_{\omega \Lambda}$ &7.0733 &7.849 &7.16446 &8.2077 \\ 
    $g_{\omega \Sigma}$ &7.0733 &7.849 &7.16446 &8.2077 \\ 
    $g_{\omega \Xi}$ &3.5366 &3.925 &3.58223 &4.1039\\ 
    \hline 
    $g_{\rho \Lambda}$ &0.0 &0.0 &0.0  &0.0\\ 
    $g_{\rho \Sigma}$ &4.10 &10.187 &9.9829 &28.3235\\ 
    $g_{\rho \Xi}$ &8.20 &10.187 &9.9829 &14.1618\\ 
    \hline 
    $g_{\phi \Lambda}$ &-6.02627 &-5.5504 &-5.0660 &5.8037\\
    $g_{\phi \Sigma}$ &-6.02627 &-5.5504 &-5.0660 &5.8037 \\
    $g_{\phi \Xi}$ &-8.9785 &-11.1008 &-10.1321 &11.6075\\
    \hline 
    $g_{\xi \Lambda}$ &1.914 &0.0 &0.0 &0.0  \\ 
    $g_{\xi \Sigma}$ &0.0 &0.0 &0.0 &0.0\\ 
    $g_{\xi \Xi}$ &0.0 &0.0 &0.0 &0.0
    \end{tabular} 
    \end{centering} 
    \caption{The parameters of the nonlinear RMF models that we consider in this work including saturation densities $n_0$, $\sigma$ meson mass values, and the various coupling constants. The GM1-Y5 Model includes the hidden strangeness isoscalar-scalar $\xi$ meson and includes $\sigma$ self interactions \cite{Oertel_2015, Glendenning_1991}. The Hornick 65 and 70 models include an additional $\omega^2\rho^2$ interaction term \cite{Hornick_2018}, and the Big-Apple model include the quartic $\omega^4$ self interaction as well \cite{Das_2021, Fattoyev_2020}. The models chosen sample a wide range of baryon-meson and meson-meson interactions as a result, allowing us to investigate possible $g-$mode dependence on interaction specific terms, but still produce stars that satisfy all current astrophysical constraints.}
    \label{tab:nonlinear_rmf_models}
\end{table}

% \begin{figure}[H]
%     \centering
%     \includegraphics[trim = 0 40 0 0, clip, 
%     scale=0.75]{Model_Images/GM1-Y5 frac_hyp.pdf}
%     \caption{Particle fraction versus total baryon number density for the GM1-Y5 model }
%     \label{fig:gm1_y5_frac}
% \end{figure}

\begin{figure}
    \centering
    \includegraphics[width = 0.47\textwidth]{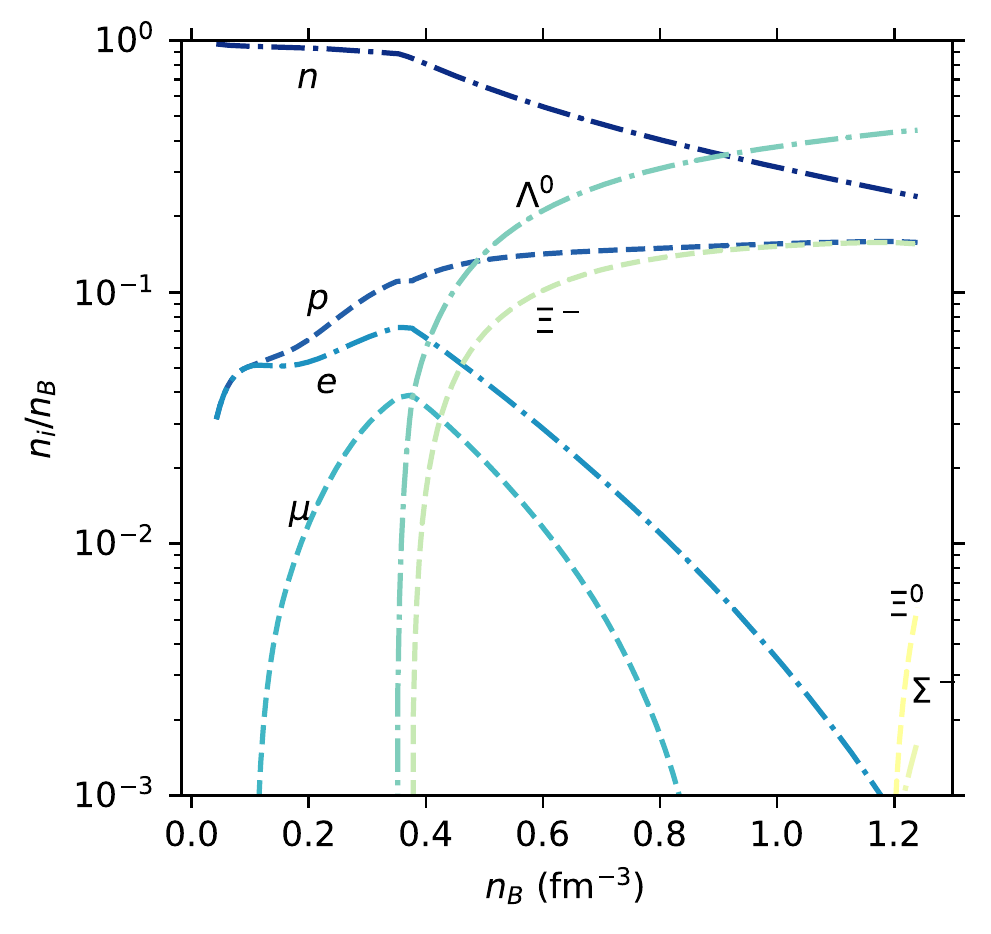}
    \caption{Particle fractions ($n,p,e,\mu, \Lambda^0, \Xi^-, \Xi^0, \Sigma^-)$ as a function of total baryon number density $n_B$ for the Big Apple EDF model including hyperons.} 
    \label{fig:big_apple_frac}
\end{figure}

To solve a particular model for neutron star matter, i.e., to obtain the particle fractions as a function of baryon density in the star as shown in Fig.\ref{fig:big_apple_frac} for the Big Apple EDF model~\footnote{The particle fractions in the other models are qualitatively similar to the Big Appled EDF results shown in Fig.\ref{fig:big_apple_frac}.}, we first derived the Euler-Lagrange equations of motion for the mesons from the Lagrangian. Then we applied our constraints of baryon number conservation, charge neutrality
\begin{align}
    n_B &= \sum_i n_i \qquad i \in \text{baryons}\\
    0 &= \sum_j q_j n_j \qquad j \in \text{baryons, leptons}
\end{align}
and imposed chemical equilibrium with respect to weak processes
%which we will repeat here for completeness
%\begin{align}
 %   \mu_i &= \mu_n - q_i \mu_e \qquad i \in \text{baryons}\\
 %   \mu_\ell &= \mu_e  
%\end{align}
%for each independent baryon or lepton in our system. 

% Each independent variable generates its own beta-equilibrium condition equation. 

As a result, for a system of $m$ mesons and $n$ baryons and leptons, we are left with $m + n + 1$ unknowns: the $m$ meson field values, the $n$ baryon and lepton fractions and the total baryon number density $n_B$. Taking $n_B$ to be our free variable, we solved for the remaining $m$+$n$ variables at that given value for $n_B$. 
One final point to consider is that for lower values of $n_B$ (which would correspond to the outer layers of the core or lower mass stars), it may not be energetically favorable for heavier particles such as hyperons to appear. Threshold conditions for the emergence of a new particle species are \cite{glendenning_1985}
\begin{align}
    \mu_n - q_b \mu_e \geq \mu_b^{(0)}
\end{align}
where 
\begin{align}
    \mu_b^{(0)} = m_i^* + \mu_i^{(m)} = m_i^* + g_{\omega i}\omega + g_{\rho i}I_{3i}\rho + g_{\phi i}\phi 
\end{align}
Whenever this condition is satisfied, we add the baryon to our system which involves updating the system of equations.

From the Lagrangian we can write down the energy momentum tensor \cite{glendenning_1985} 
\begin{align}
    \mathcal{T}^{\mu\nu} &= \sum_n \frac{\partial \mathcal{L}}{\partial(\partial_\mu \phi_n)}\partial_\nu \phi_n - g_{\mu\nu}\mathcal{L}
\end{align}
which yields the energy density $\varepsilon$ and pressure $p$. In the mean field approximation, the energy density is then
\begin{equation} 
\begin{aligned}
    \varepsilon &= \frac{1}{2}m_\sigma^2 \sigma^2 + 
    \frac{1}{2}m_\rho^2 \rho^2 + 
    \frac{1}{2}m_\phi^2 \phi^2\\
    &\qquad + \frac{1}{2}m_\xi^2 \xi^2 + 
    \frac{1}{2}m_\delta^2 \delta^2 + \frac{1}{3}bm_N (g_{\sigma N}\sigma)^3\\
    &\qquad + \frac{1}{4}c(g_{\sigma N}\sigma)^3 + 3\Lambda_\omega g_{\omega N}^2 g_{\rho N}^2 \omega^2 \rho^2 \\
    &\qquad + \sum_{i \in B}\frac{2J_i + 1}{2\pi^2} \int_0^{k_{F_i}}\sqrt{k^2 + {m_i^*}^2}\,k^2\,dk\\
    &\qquad + \sum_{\ell}\frac{2J_i + 1}{2\pi^2} \int_0^{k_{F_\ell}}\sqrt{k^2 + {m_i^*}^2}\,k^2\,dk \label{eqn:energy_density}
\end{aligned}
\end{equation}
with the integrals evaluating to
\begin{equation} 
\begin{aligned}
    \int_0^{k_{F_i}} \sqrt{k^2 + {m_i^*}^2}\,k^2\,dk = \frac{1}{4}\bigg[k_{F_i}(k_{F_i}^2 + {m_i^*}^2)^{3/2}\\
    + k_{F_i}^3\sqrt{k_{F_i}^2 + {m_i^*}^2} - {m_i^*}^4\ln \frac{k_{F_i} + \sqrt{k_{F_i}^2 + {m_i^*}^2}}{m_i^*}\bigg] 
\end{aligned}
\end{equation} 

The pressure follows from the relation \cite{Pradhan_2022, Hornick_2018}. 
% \begin{align}
%     \varepsilon &= \varepsilon_\text{meson} + \varepsilon_\text{B} + \varepsilon_\text{lepton}\\
%     &= \sum_{K} \frac{1}{2} m_K^2 K^2 \qquad K \in \sigma,\omega,\rho,\dots\\
%     &\qquad + 
% \end{align}
% and 
% \begin{align}
%     P = 
% \end{align}
\begin{align}
    P &= \sum_i \mu_i n_i - \varepsilon \label{eqn:thermo} 
\end{align}
With $p$ and $\varepsilon$, we can determine the equilibrium sound speed from
\begin{align}
    c_e^2 := \frac{dP}{d\varepsilon} = \frac{dP}{d n_B} \frac{dn_B}{d\varepsilon} = \frac{dP}{dn_B} \frac{1}{d\varepsilon/dn_B }
\end{align}

Next, we need to consider how to generate the hyperonic coupling constants. Starting with the nucleon-meson couplings that are chosen to satisfy saturation density properties~\cite{Han_2019, Oertel_2017}, we can then generate hyperonic couplings via relationships similar to that expressed in Eq.(\ref{expr_1}), with the full list of relationships given in \cite{Oertel_2015}. 
\begin{align}
    \frac{g_{\omega \Lambda}}{g_{\omega N}} &= 
    \frac{1 - \frac{2z}{\sqrt{3}}(1- \alpha)\tan\theta}{1 - \frac{z}{\sqrt{3}}(1-4\alpha)\tan\theta} \label{expr_1}
\end{align}
If we take the ideal mixing limit, $\alpha = 1$, $z = 1/\sqrt{6}$, and $\tan\theta = 1/\sqrt{2}$, the relations respect SU(6) symmetry \cite{Miyatsu_2013,  schaffner_multiply_1994}.
    \begin{align} 
    g_{\omega\Lambda} &= g_{\omega \Sigma} = 2g_{\omega \Xi} = \frac{2}{3}g_{\omega N}\\
    g_{\rho \Lambda} &= 0 \qquad g_{\rho \Sigma} = 2g_{\omega \Xi} = 2g_{\omega N}\\
    g_{\phi N} &= 0 \qquad 2g_{\phi \Lambda} = 2g_{\phi \Sigma} = g_{\phi \Xi} = \frac{2\sqrt{2}}{3}g_{\omega N}
\end{align}
For most of our models, we generate the hyperonic-vector meson coupling constants using these SU(6) relations~\footnote{The GM1-Y5 model takes $z = 0.2$ rather than $z = 1/\sqrt{6}$ \cite{Oertel_2015}}. The coupling constants for scalar fields that couple to mass, that is, the scalar sigma meson, are determined by fitting to hyperonic optical potentials via the following equation \cite{Miyatsu_2013, Thapa_2021}
\begin{align}
    U_Y^{(N)} = -  g_{\sigma Y} \sigma_0 + g_{\omega Y}\omega_0 \label{eqn:general}
\end{align}
where $U_Y^{(N)}$ is the corresponding hyperon optical potential, $\sigma_0$, $\omega_0$, are the saturation density values for the $\sigma$, $\omega$ mesons which can be found by solving the standard $npe$ case first, and $g_{\omega Y}$ is the omega-hyperon coupling which can be determined using SU(6) relations described previously. For the values of the hyperon potentials at saturation density, we use the most commonly accepted values for $U_\Lambda^{(N)} = - 30$\,MeV and $U_\Sigma^{(N)} = 30$\,MeV \cite{Tu_2022, Pradhan_2022, PhysRevC.62.034311, Thapa_2021, Rather_2021}. Although $U_\Xi^{(N)}$ is known to be attractive, its precise value at saturation is not well constrained \cite{Pradhan_2022, PhysRevC.62.034311}. For this work, we take it to be $U_\Xi^{(N)} = -14,-15$\,MeV in accordance with currently used models \cite{Tu_2022, Thapa_2021, Rather_2021}. The hyperonic coupling constants for the strange scalar mesons $g_{\xi Y}$ and $g_{\delta Y}$ can be found by fitting to a more general version of eqn. (\ref{eqn:general}) 
\begin{align}
    U_j^{(k)}(n_k) = m_j^* - m_j + \mu_j - \mu_j^*
\end{align}
at densities above saturation when strange degrees of freedom emerge, as a second step after fitting to saturation
\cite{Oertel_2015}. However, for the work done here, $g_{\xi Y}$ is only used for GM1-Y5 with values taken as specified in \cite{Oertel_2015} and $g_{\delta Y}$ is not calculated as there is no $\delta$ meson dependence in the models that we choose to consider.

\subsection{Density Dependent RMF Models (DDRMF)}
\begin{figure*}
     \centering
     \begin{subfigure}[b]{0.32\textwidth}
         \centering
         \includegraphics[width=\textwidth]{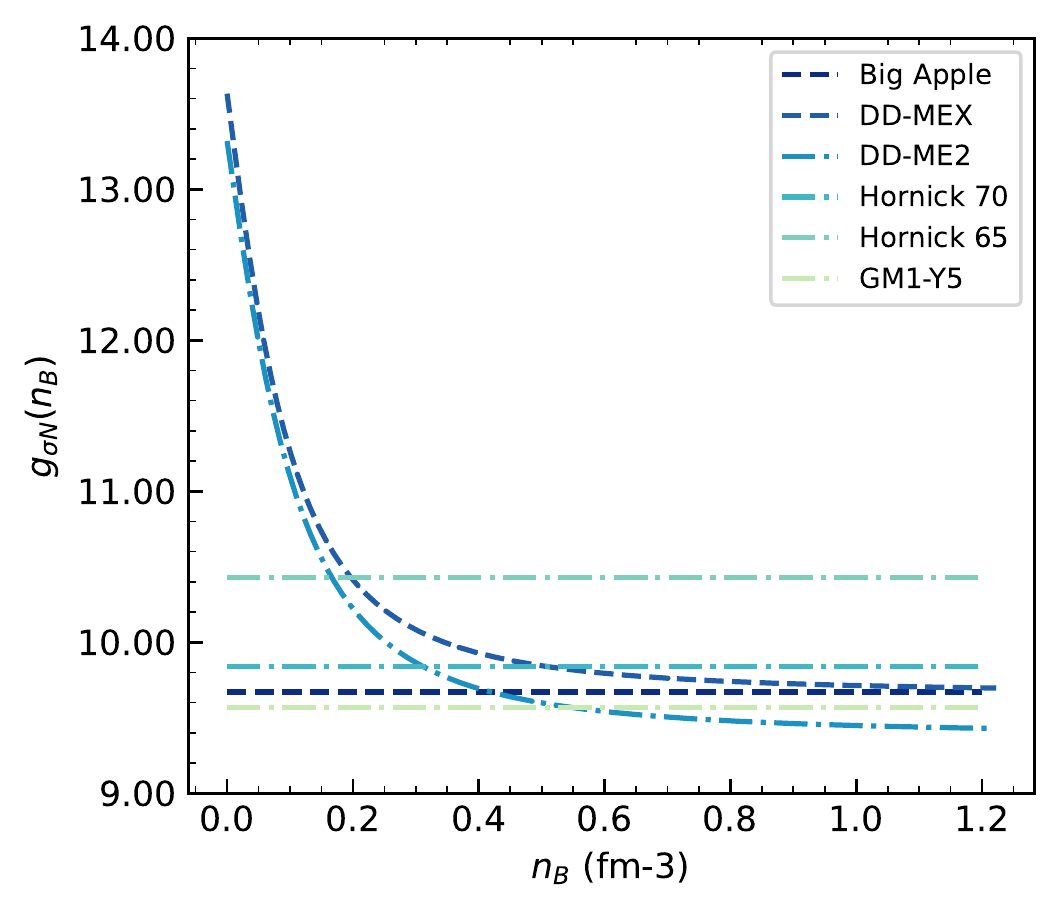}
         \label{fig:g_sigma_n_dd_rmf}
     \end{subfigure}
     \hfill
     \begin{subfigure}[b]{0.32\textwidth}
         \centering
         \includegraphics[width=\textwidth]{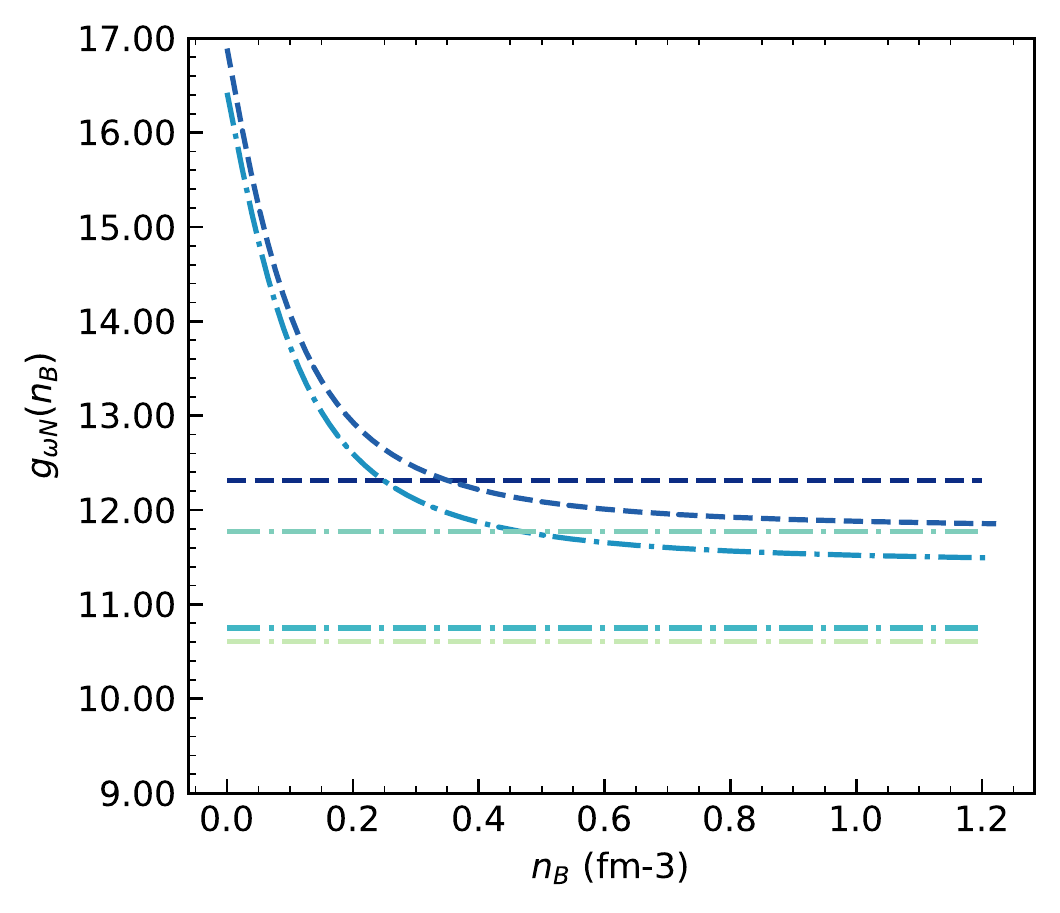}
         \label{fig:g_omega_n_dd_rmf}
     \end{subfigure}
          \hfill
     \begin{subfigure}[b]{0.32\textwidth}
         \centering
         \includegraphics[width=\textwidth]{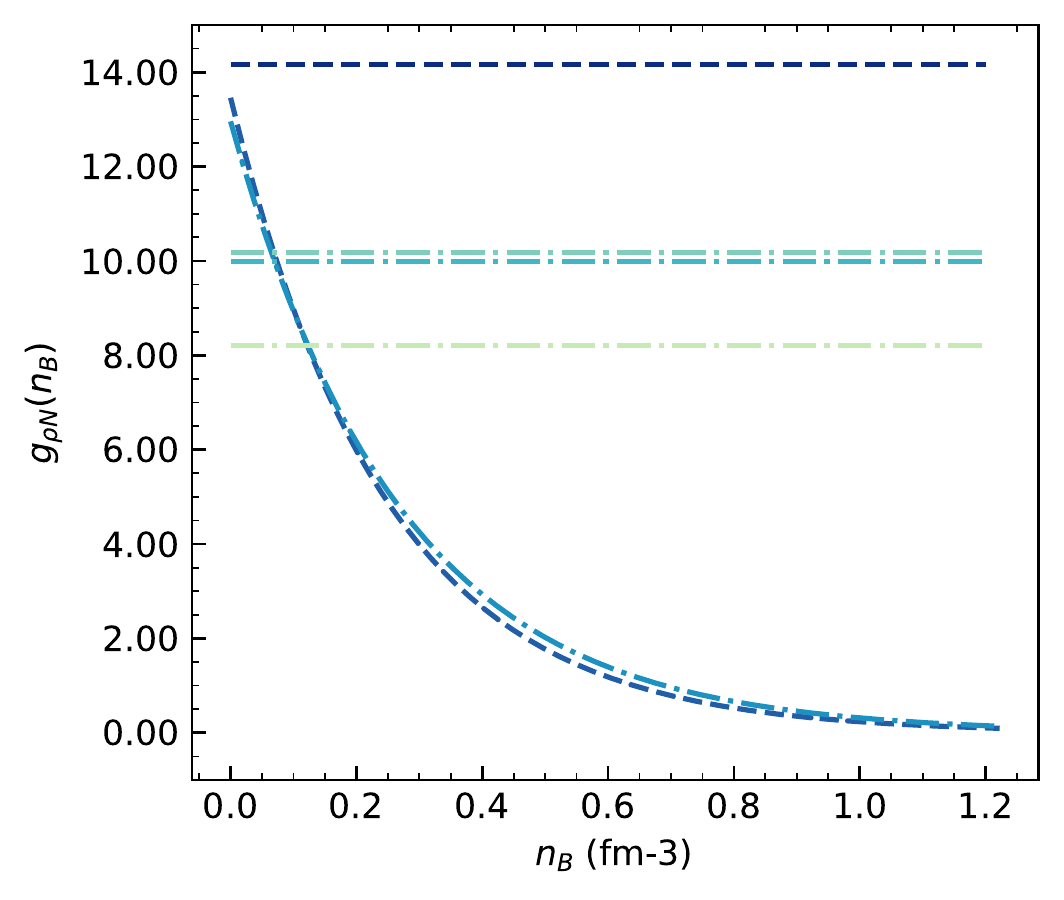}
         \label{fig:g_rho_n_dd_rmf}
     \end{subfigure}
     \caption{Density dependence of coupling constants $g_{\sigma N}$, $g_{\omega N}$, and $g_{\rho N}$  for the various DDRMF models. Plotted for reference are the constant coupling constants from the NLRMF models. We see that for large $n_B$ that $g_{\sigma N}$ and $g_{\omega N}$ behave similarly to those in the NLRMF models whereas $g_{\rho N}$ decreases towards zero indicating that in the DDRMF models  isospin interactions vanish as $n_B\approx 8n_0$.} 
\end{figure*}
A related class of models to the NLRMF type of models is the density dependent relativistic mean field model (DDRMF) where the baryon-meson coupling constants are allowed to vary with baryon number density $n_B$ rather than remain constant throughout the entire range of densities. The coupling constants become density dependent and typically take on the following forms for the $\sigma, \omega, \phi$ mesons
\begin{align}
    g_i(n_B) 
    &=g_i(n_0) \cdot a_i \frac{1 + b_i(n_B/n_0 + d_i)^2}{1 + c_i(n_B/n_0 + d_i)^2} \label{eqn:dd_rmf_gi}
\end{align}
and 
\begin{align}
    g_\rho(n_B) &= g_{\rho}(n_0) a_\rho \exp\left(-a_\rho\left(\frac{n_B}{n_0}-1\right)\right) \label{eqn:dd_rmf_grho}
\end{align}
for the $\rho$ meson where $g_i(n_0)$ is the coupling constant at saturation and $a_i, b_i, c_i, d_i$ are additional parameters that determine the evolution of the coupling constants for the models \cite{Tu_2022, Huang_2020, Rather_2021, Thapa_2021, Kl_hn_2006}.

There is an additional term added to the chemical potential $\mu_i$ called the rearrangement term $\Sigma^r$ for thermodynamic reasons \cite{Tu_2022, Thapa_2021, Rather_2021} 
\begin{align}
\label{rearr}
    \mu_i^{(b)} &= E_{F_i}^* + \mu_i^{(m)} + \Sigma^r(n_B)
\end{align}
where for an interaction Lagrangian which includes the scalar-isoscalar $\sigma$, vector-isoscalar $\omega$, vector-isovector $\rho$ and hidden-strangeness vector-isoscalar $\phi$ mesons 
\begin{align}
    \mathcal{L}_\text{int} &= - \sum_i \bar{\psi}_i(\gamma_0 \mu_i^* - m_i^*)\psi_i 
\end{align}
with $\mu_i^* = \mu_i - g_{\omega i}\omega - I_{3i} g_{\rho i}\rho - g_{\phi} - \Sigma^r$ and $m_i^* = m_i - g_{\sigma i}\sigma$, $\Sigma^r$ takes the form \cite{Huang_2020}
\begin{equation} 
\begin{aligned}
    \Sigma^r(n_B) &= \sum_i \bigg[-\frac{\partial g_\sigma(n_B)}{\partial n_B}\sigma n_i^s + \frac{\partial g_{\omega i}(n_B)}{\partial n_B} \omega n_i\\
    &\qquad 
    + \frac{\partial g_{\rho i}}{\partial n_B}\tau^3_i\rho n_i
    + \frac{\partial g_{\phi i}}{\partial n_B}\phi n_i 
    \bigg] \,.
\end{aligned}
\end{equation} 

This rearrangement term contributes to the expression for pressure $p$ , though not the energy density $\varepsilon$ which takes on the same form as Eq.(\ref{eqn:energy_density}) which allows us to determine $p$ through the thermodynamic relationship with $\varepsilon$ given in Eq.(\ref{eqn:thermo}).

\begin{table}[h!]
    \centering
    \begin{tabular}{ccccccc}
    \hline 
    \hline 
      Model   & DD-MEX  &DD-ME2  \\
    \hline 
    \hline 
    $n_0$ (fm$^{-3}$)  & 0.152 & 0.152\\
    $m_\sigma$ (MeV) &547.333  &550.124 \\ 
    \hline 
     $g_{\sigma N}(n_0)$ &10.707  & 10.540\\ 
     $g_{\omega N}(n_0)$ &13.339  &13.019\\
     $g_{\rho N} (n_0)$ &7.238     &7.367\\
     \hline 
     $a_\sigma$ &1.397    & 1.388\\ 
     $b_\sigma$ &1.335   &1.094\\ 
     $c_\sigma$ &2.067 &1.706 \\ 
     $d_\sigma$ &0.402 &0.442 \\ 
     \hline 
     $a_\omega$ &1.394  &1.389 \\
     $b_\omega$ &1.019  &0.924\\ 
     $c_\omega$ &1.606  &1.462\\ 
     $d_\omega$ &0.456  &0.478\\ 
     \hline 
     $a_\rho$ &0.620  &0.565\\ 
     \hline \\ 
    \end{tabular}
    \caption{Different DDRMF Models chosen for this work and their parameters including saturation density $n_0$, non-strange meson coupling constants, and density dependent specific parameters \cite{Rather_2021, Thapa_2021}. In particular, $g_{\sigma N}(n_0), g_{\omega N}(n_0), g_{\rho N}(n_0)$ refer to the values of the coupling constants at saturation and $a_i, b_i, c_i, d_i$ determine the dependence of the coupling constants on total baryon number density $n_B$ as given by eqns. (\ref{eqn:dd_rmf_gi}) and (\ref{eqn:dd_rmf_grho}).}
    \label{tab:dd_rmf_parameters}
\end{table}
\begin{table}
    \centering
    \begin{tabular}{ccccccc}
    \hline 
    \hline 
    Model     &DD-MEX  &DD-ME2  \\
    \hline 
    \hline 
    $g_{\sigma \Lambda}(n_0)$     &6.613  &6.535 \\
    $g_{\omega \Lambda}(n_0)$   &8.893  &8.679 \\ 
    $g_{\rho \Lambda}(n_0)$ &0.0  &0.0 \\ 
    $g_{\phi \Lambda}(n_0)$ &6.288  &6.137 \\
    \hline 
    $g_{\sigma \Sigma}(n_0)$ &5.0834  &4.962 \\ 
    $g_{\omega \Sigma}(n_0)$ &8.893  &8.679 \\ 
    $g_{\rho \Sigma}(n_0)$ &14.476 &14.734 \\ 
    $g_{\phi \Sigma}(n_0)$ &6.288 &6.137  \\ 
    \hline 
    $g_{\sigma \Xi}(n_0)$ &3.3319  &3.320 \\ 
    $g_{\omega \Xi}(n_0)$ &4.446  &4.340 \\
    $g_{\rho \Xi}(n_0)$ &7.238 &7.367 \\ 
    $g_{\phi \Xi}(n_0)$ &12.576 &12.274 \\
    \hline 
    \end{tabular}
    \caption{Hyperon couplings using SU(6) symmetry arguments as determined by fitting saturation density coupling constants and fields to to Eq. \ref{eqn:hyperon_fit_ddrmf} using the following values for the hyperon optical potentials $U_\Lambda^N = - 30$\,MeV, $U_\Sigma^N = 30$\,MeV, $U_\Xi^N = -14$\,MeV \cite{Tu_2022, Pradhan_2022, PhysRevC.62.034311, Thapa_2021, Rather_2021}.}
    \label{tab:dd_rmf_hyp_couplings}
\end{table}
\begin{figure*}
    \centering
    \begin{subfigure}[b]{0.47\textwidth}
        \includegraphics[width = \textwidth]{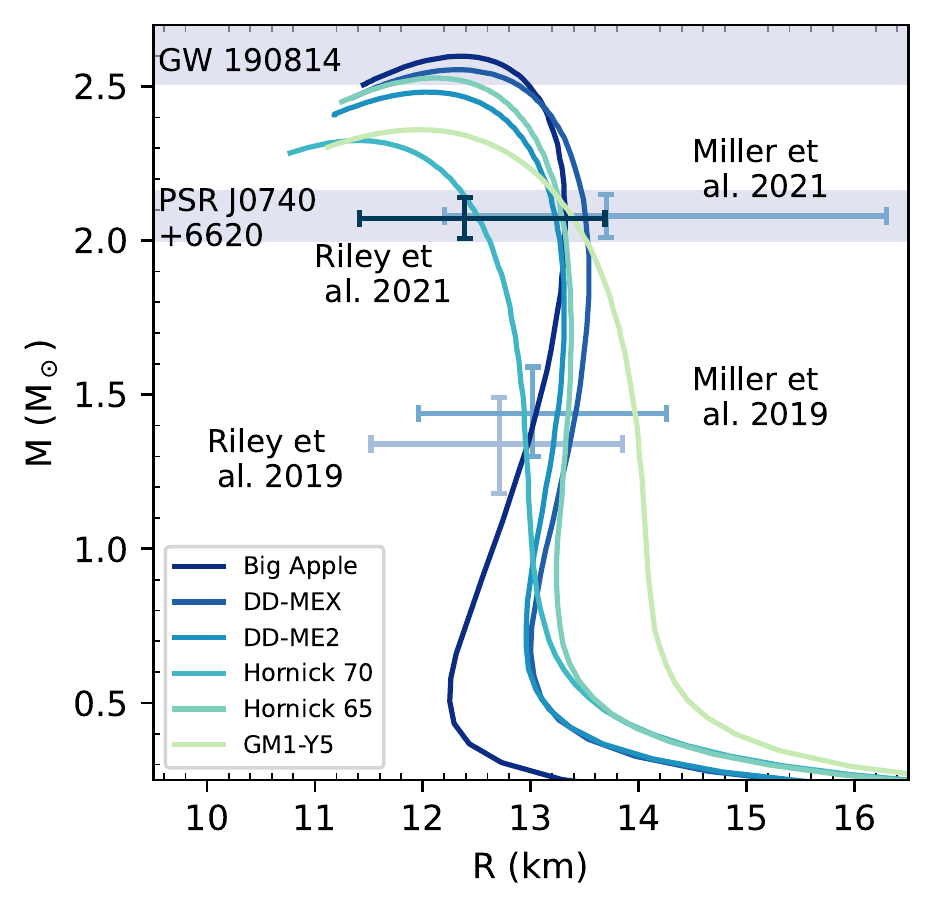}
        \caption{$npe\mu$} 
        \label{fig:mass_radius_npemu} 
    \end{subfigure}
    \hfill 
    \begin{subfigure}[b]{0.47\textwidth}
          \includegraphics[width = \textwidth]{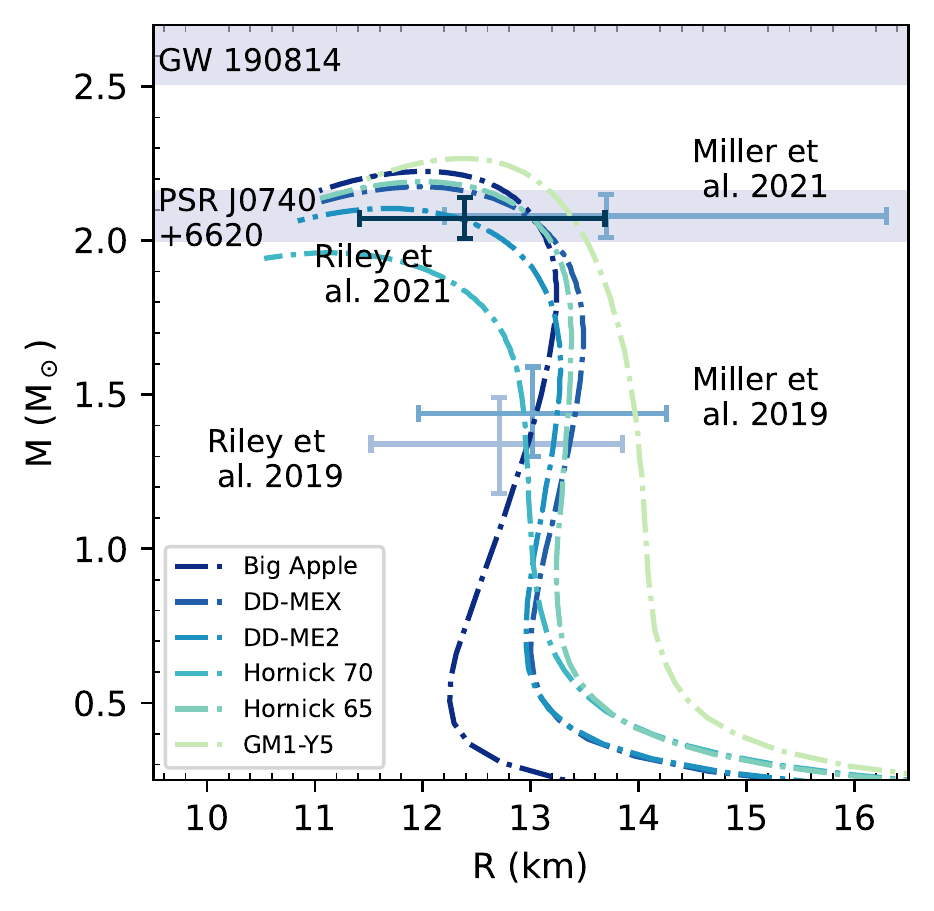}
        \caption{$npe\mu Y$} 
        \label{fig:mass_radius_npemuy} 
    \end{subfigure}
    \caption{Mass radius relations for the various models used in this work with the $npe\mu$ and $npe\mu Y$ compositions on the left and right respectively. The astrophysical constraints of maximum masses from PSR J0740+6620 and the secondary object of GW 190814 are likewise plotted here in the light blue areas \cite{Miller_2021, Abbott_2020}. Additionally plotted are the Neutron Interior Composition Explorer (NICER) constraints on the mass-radius of PSR J0030+0451 from Riley et al. 2019 ($1.34_{-0.16}^{+0.15}\,M_\odot$ and $12.71^{1.14}_{-1.19}\,\text{km}$) and Miller et al. 2019 ($1.44^{+0.15}_{-0.14}\,M_\odot$ and $13.02_{-1.06}^{+1.24}\,\text{km}$) \cite{Riley_2019, Miller_2019} as well as for PSR J0740+6620 with values of ($2.072_{-0.066}^{+0.067}\,M_\odot$ and $12.39_{-0.98}^{+1.30}\,\text{km}$) and ($2.08_{-0.07}^{+0.08}\,M_\odot$ and  $13.7_{-1.5}^{+2.6}\,\text{km}$) \cite{Riley_2021, Miller_2021}}
    \label{fig:mass_radius}
\end{figure*}
To get the hyperon coupling constants, we can employ the same SU(6) symmetry scheme as mentioned in Sec \ref{sec:NLRMF}. The equation relating the hyperon optical potentials to the $g_{\sigma Y}$ coupling constants is modified to include the $\Sigma^r$ term
\begin{align}
    U_Y^N &= g_{\omega Y}\omega_0 - g_{\sigma Y}\sigma_0 + \Sigma^r(n_B) \label{eqn:hyperon_fit_ddrmf}
\end{align}
which for the previously mentioned hyperon optical potentials of $U_\Lambda^N = - 30$ MeV, $U_\Sigma^N = 30$ MeV, $U_\Xi^N = -14$ MeV yield the relations $g_{\sigma \Lambda} = 0.6105\, g_{\sigma N}$, $g_{\sigma \Sigma} = 0.4426\,g_{\sigma N}$, and $g_{\sigma \Xi} = 0.3024\,g_{\sigma N}$ \cite{Rather_2021}. The corresponding hyperon coupling constants are listed in Tab. \ref{tab:dd_rmf_hyp_couplings}. 

Ultimately, we choose the following density dependent RMF models: DD-MEX \cite{Taninah_2020} and DD-ME2 \cite{Tu_2022} as they produce stars with mass-radius curves and tidal deformabilities in agreement with current astrophysical constraints from NICER and GW170817. Their relevant parameters including coupling constants are listed in Tab. \ref{tab:dd_rmf_parameters}. For these models, $U_{NL}$ is effectively zero, that is, there is no non-linear meson-meson interactions unlike in the non-linear RMF models. These models are likewise fit to the following saturation parameters: $E_0 = - 16.14$ MeV, $K_0 = 267.059$ and $250.89$ MeV respectively \cite{Thapa_2021}. 

\begin{figure}[h!]
    \centering
    \includegraphics[scale=0.85]{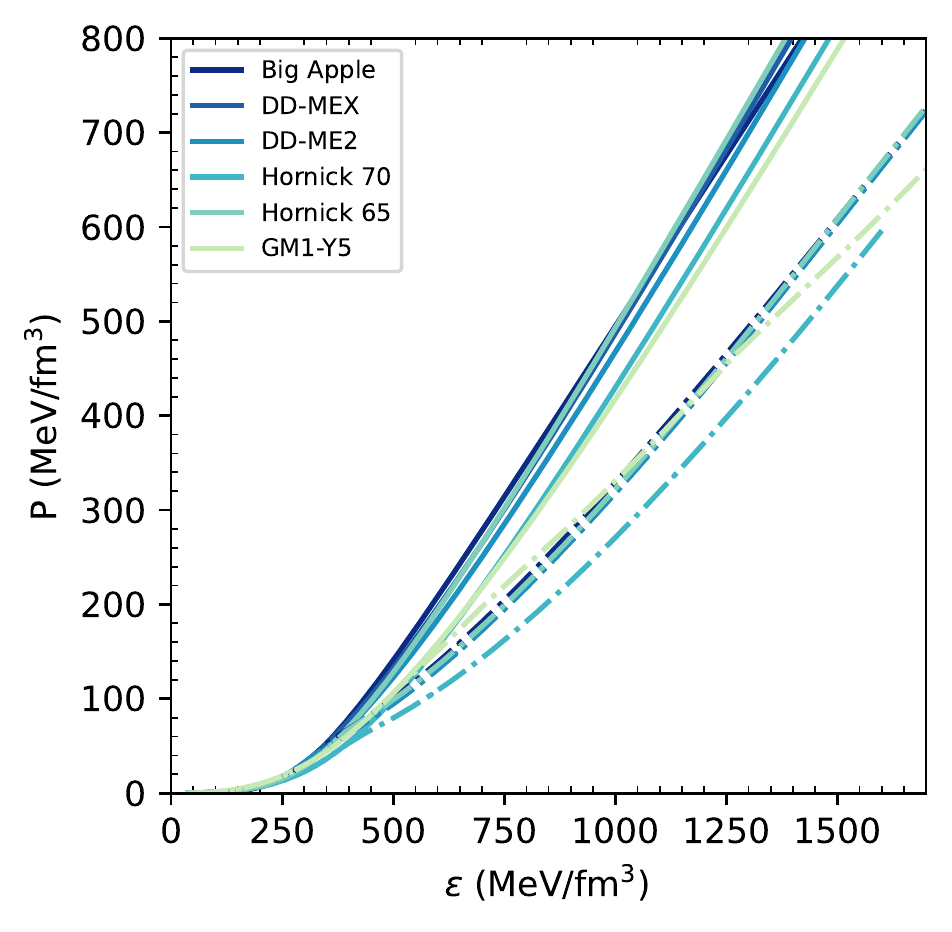}
    \caption{The equation of state for the various models considered in this work with $npe\mu$ matter (solid line) and $npe\mu$Y matter (dot-dashed line). The onset of hyperons leads to a characteristic softening of the equation of state \cite{Bedaque_2015, 2017hspp.confj1002B} .}
    \label{fig:eos}
\end{figure}

\begin{figure}
    \centering
    \includegraphics[scale=0.80]{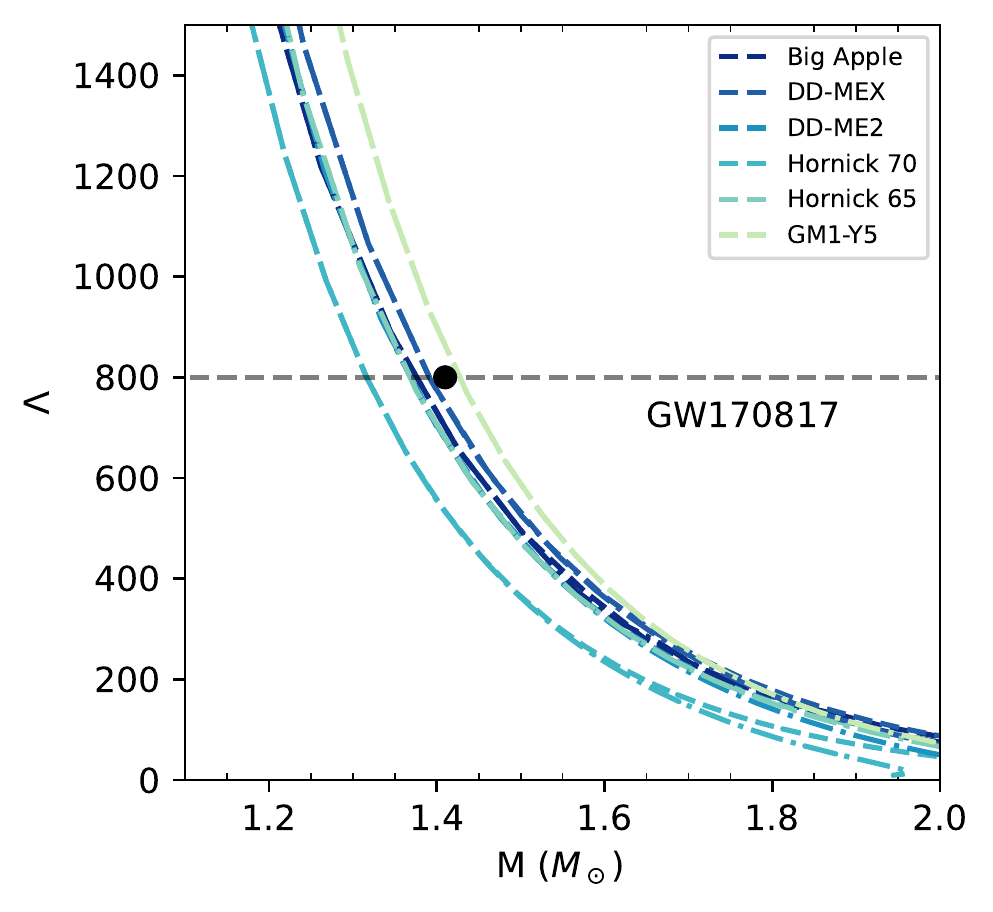}
    \caption{Tidal Deformability plotted against stellar mass for our various models. We see that all save for GM1-Y5 are safely below the $\Lambda \leq 800$ constraint from GW170817}
    \label{fig:tidal_deformability_linear}
\end{figure}

\subsection{Equilibrium Structure} 

The pressure $p$ and $\varepsilon$ tabulated against total baryon number density give us a parameteric equation of state, Fig. \ref{fig:eos} which then determines the macroscopic properties like mass and radius of the star from the Tolman-Oppenheimer-Volkov (TOV) equations eqns. (\ref{TOV1}) and (\ref{TOV2}) for a static, spherically symmetric star in hydrostatic equilibrium. Fig. \ref{fig:mass_radius_npemu}, \ref{fig:mass_radius_npemuy} and Fig. \ref{fig:tidal_deformability_linear}
show the corresponding mass-radius plots and tidal deformability with observational constraints (as error bars). The models we use satisfy current observational astrophysical constraints.

\begin{align}
\label{TOV1}
    \frac{dp}{dr} &= - \frac{G m(r)\varepsilon(r)}{r^2}\frac{\left[1 + \dfrac{p(r)}{\varepsilon(r)}\right]\left[1 + \dfrac{4\pi r^3 p(r)}{m(r)}\right]}{1- \dfrac{2GM(r)}{r}}\\
    \frac{dm}{dr} &= 4\pi \varepsilon(r) r^2
    \label{TOV2}
\end{align}

\section{Adiabatic Sound Speed via Sound Speed Difference \label{sec:adiabatic_sound_speed}}
Having established our working models for the stellar structure and composition, we turn now to the calculation of the adiabatic sound speed squared $c_s^2$, or equivalently, the sound speed difference (since $c_e^2$ is easily obtained from the EoS) by using Eq.(\ref{cs2_ce2}). Starting from Eq.(\ref{rearr}), the partial derivative of the baryonic chemical potential \footnote{The partial derivatives for leptons can be obtained from their relativistic dispersion relation.
\begin{align}
    \frac{\partial \mu_\ell}{\partial n_B}\bigg|_\chi &= \frac{\pi^2 x_\ell}{k_{F_\ell} E_{F_\ell}} \qquad x_\ell := \frac{n_\ell}{n_B}
\end{align}} is
\begin{align}
    \frac{\partial\mu_i^{(B)}}{\partial n_B}\bigg|_\chi = 
    \frac{\partial E_{F_i}^*}{\partial n_B}\bigg|_\chi + \frac{\partial \mu_i^{(m)}}{\partial n_B}\bigg|_\chi + \frac{\partial \Sigma^r}{\partial n_B}\bigg|_\chi
\end{align}

We discuss each of these contributions in turn, noting that the effective energy $E_{F_i}^*$ will only couple to the scalar mesons and $\mu_i^{(m)}$ will only couple to the vector mesons. 

\subsection{Partial Derivative of the Effective Energy ($E_{F_i}^*$) \label{sec:effective_energy}}
%In order to calculate the partial derivative of $E_{F_i}^*$, we start by differentiating eqn. \ref{eqn:effective_energy} with respect to $n_B$ and arrive at a system of linear equations in which the $\partial E_{F_i}^*/\partial n_B$ and the partial derivatives of the mesonic fields such as $\partial \sigma/\partial n_B$ are solutions due to the coupling of $E_{F_i}^*$ to the scalar mesons. 

Through Eq. (\ref{eqn:effective_energy}), $E_{F_i}^*$ depends on each of the scalar meson fields (say, $m$ in number) $\sigma,\delta,\xi$ through the effective mass term $m_i^* = m_i - g_{\sigma i}\sigma - g_{\xi i}\xi - I_{3i}g_{\delta i}\delta$ for the NLRMF and DDRMF models. To determine $\partial E_{F_i}^*/\partial n_B$, we would need to determine the partial derivatives
$\partial \sigma/\partial n_B, \partial \xi/\partial n_B$ and $\partial \delta/\partial n_B$ as well.  
%However, as given in Appendix \ref{sec:mesonic_mean_field}, the mesonic mean field equations of motion demonstrate the coupling of the scalar mesons both to themselves and to the scalar densities of the various baryons in the system $n_i^s$ which at zero temperature is given by \cite{glendenning_1985}
First, for each of the baryons (say, $b$ in number), we have equations for the scalar density
\begin{align}
    n_i^s = \langle \bar{\psi}_i\psi_i \rangle
    &= \frac{1}{\pi^2}
    \int_0^{k_{F_i}} \frac{m_i^*}{E_{F_i}^*}k^2\,dk \\ \nonumber
    &= \frac{m_i^*}{2\pi^2} 
    \left[k_{F_i}E_{F_i}^* - {m_i^*}^2\ln \frac{k_{F_i} +E_{F_i}^*}{m_i^*}\right] 
    \label{ns}
\end{align}
providing additional relations between $E_{F_i}^*$ and the meson fields. As a result, after differentiating both sides of eqns. (\ref{eqn:effective_energy}) and eqns. (\ref{eqn:sigma_eom} - \ref{eqn:delta_eom}), we arrive at a system of $m+b$ equations that are linear in the quantities of interest, and in particular, can be solved for $\partial E_{F_i}^*/\partial n_B$. As a concrete illustration, in the NLRMF model, starting with eqns. (\ref{eqn:effective_energy}) for $E_{F_i}^*$, we arrive at 
\begin{align}
    \frac{\partial E_{F_i}^*}{\partial n_B}\bigg|_\chi = 
    \frac{k_{F_i}}{E_{F_i}^*} - \frac{ m_i^*}{E_{F_i}^*}\frac{\partial m_i^*}{\partial n_B}\bigg|_\chi 
\end{align}
where 
\begin{align}
    \frac{\partial m_i^*}{\partial n_B}\bigg|_\chi  &= - g_{\sigma i} 
    \frac{\partial \sigma}{\partial n_B}\bigg|_\chi  - g_{\xi i}\frac{\partial \xi}{\partial n_B}\bigg|_\chi 
    - I_{3i} g_{\delta i}\frac{\partial \delta}{\partial n_B}\bigg|_\chi 
\end{align}
As for each baryon, there is an associated $E_{F_i}^*$, each contributes for a total of $b$ of these equations. Next, from the equation of motion for the the $\sigma$ meson in particular (eqn. (\ref{eqn:sigma_eom})), after differentiating, we see it likewise depends on $\partial E_{F_i}^*/\partial n_B$ for each baryon
\begin{align} 
    \frac{\partial \sigma}{\partial n_B}\bigg|_\chi 
    \left(m_\sigma^2 + \frac{\partial^2 U}{\partial \sigma^2}\right) &= 
    \sum_j g_{\sigma j} 
    \frac{\partial n_j^s}{\partial n_B}\bigg|_\chi  \label{eqn:orig}
\end{align}
with $U = \frac{1}{3} b m_{N}(g_{\sigma N}\sigma)^3 + \frac{1}{4}c (g_{\sigma N}\sigma)^4$ and where
\begin{align}
     \frac{\partial n_j^s}{\partial n_B}\bigg|_\chi &=  
     \frac{\partial}{\partial n_B}
     \frac{m_j^*}{2\pi^2}
     \left[k_{F_j}E_{F_j}^* - {m_j^*}^2\ln \frac{k_{F_j} + E_{F_j}}{m_j^*}\right]
\end{align}
which after expanding, evaluating, and re-inserting into eqn. (\ref{eqn:orig}) leads us to eqn. (\ref{eqn:d_sigma_eom}) where the relationship between $\partial E_{F_i}^*/\partial n_B$, $\partial \sigma/\partial n_B$, $\partial \xi/\partial n_B$, and $\partial \delta/\partial n_B$ is more explicit.

\begin{widetext} 
\begin{equation} 
\begin{aligned}
    0 &=  -
         \frac{\partial \sigma}{\partial n_B}\bigg|_\chi 
    \left(m_\sigma^2 + \frac{\partial^2 U}{\partial \sigma^2}\right) - 
    \sum_i 
    \frac{g_{\sigma i} }{2 \pi^2} 
    \left(g_{\sigma i} 
    \frac{\partial \sigma}{\partial n_B}\bigg|_\chi  + g_{\xi i}\frac{\partial \xi}{\partial n_B}\bigg|_\chi 
    + I_{3i}g_{\delta i}\frac{\partial \delta}{\partial n_B}\bigg|_\chi \right)
   \frac{n_i^s}{m_i^*}\\
    &\quad + \sum_i 
    g_{\sigma i}
    \frac{m_i^*}{2\pi^2}
    \bigg[
    \frac{\pi^2 x_i}{k_{F_i}^2}  E_{F_i}^* 
    + k_{F_i}\frac{\partial E_{F_i}^*}{\partial n_B}\bigg|_\chi \bigg]\\
    &\quad  + \sum_i g_{\sigma i} \frac{m_i^*}{2\pi^2}
    \bigg[
    2g_{\sigma i}m_i^* \left(g_{\sigma i} 
    \frac{\partial \sigma}{\partial n_B}\bigg|_\chi  + g_{\xi i}\frac{\partial \xi}{\partial n_B}\bigg|_\chi 
    + I_{3i}g_{\delta i}\frac{\partial \delta}{\partial n_B}\bigg|_\chi\right) 
    \ln\frac{k_{F_i} + E_{F_i}^*}{m_i^*}\\
    &\qquad \quad \quad 
    - {m_i^*}^2
    \left[\frac{\dfrac{\pi^2 x_i}{k_{F_i}^2} + \dfrac{\partial E_{F_i}^*}{\partial n_B}\bigg|_\chi }{k_{F_i} + E_{F_i}^*} +
        \frac{1}{m_i^*}\left(g_{\sigma i} 
    \frac{\partial \sigma}{\partial n_B}\bigg|_\chi  + g_{\xi i}\frac{\partial \xi}{\partial n_B}\bigg|_\chi 
    + I_{3i}g_{\delta i}\frac{\partial \delta}{\partial n_B}\bigg|_\chi\right)\right]\bigg] \label{eqn:d_sigma_eom}
\end{aligned}
\end{equation} 
\end{widetext}

Similar equations appear when we repeat this procedure for the remaining scalar mesons, contributing a total of $m$ equations to the system. The required derivatives are solved for using standard numerical methods for a linear system of coupled equations. 
\begin{figure*}
    \centering
     \begin{subfigure}[b]{0.47\textwidth}
         \centering
         \includegraphics[width=\textwidth]{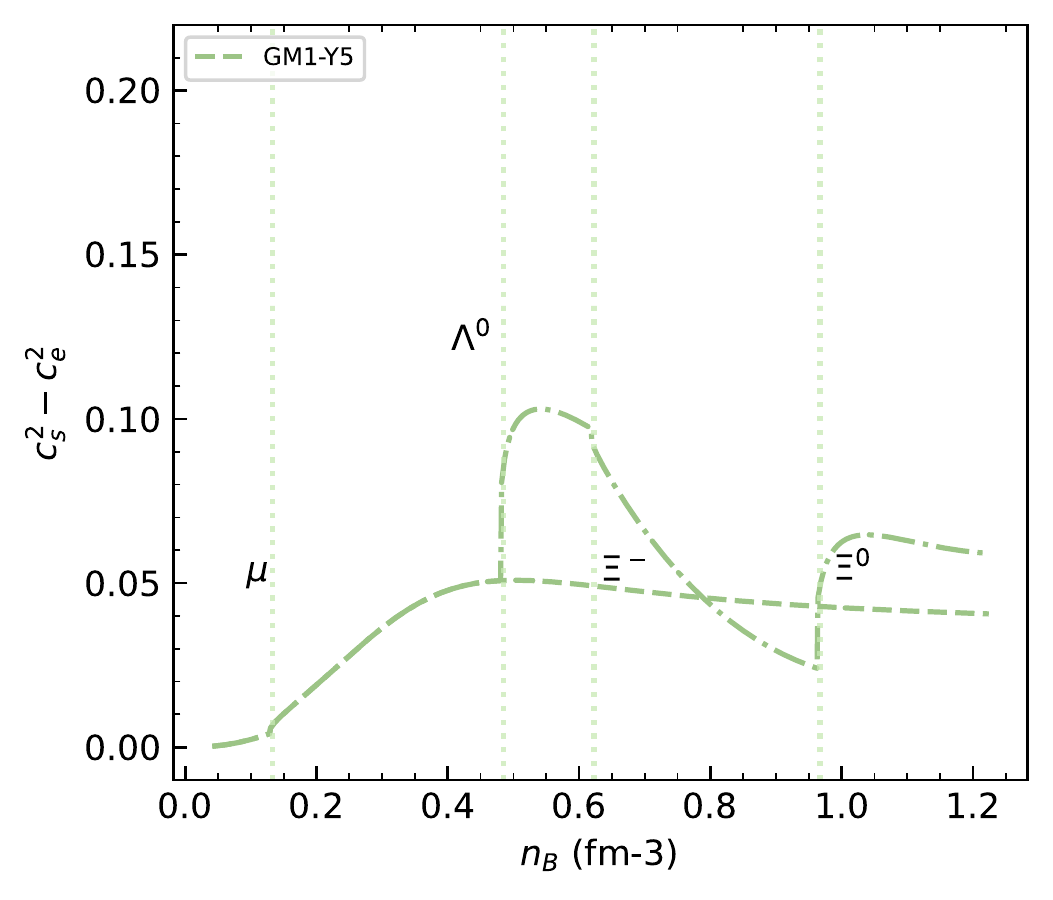}
         \label{fig:gm1_y5_sound_speed}
     \end{subfigure}
     \hfill
     \begin{subfigure}[b]{0.47\textwidth}
         \centering
         \includegraphics[width=\textwidth]{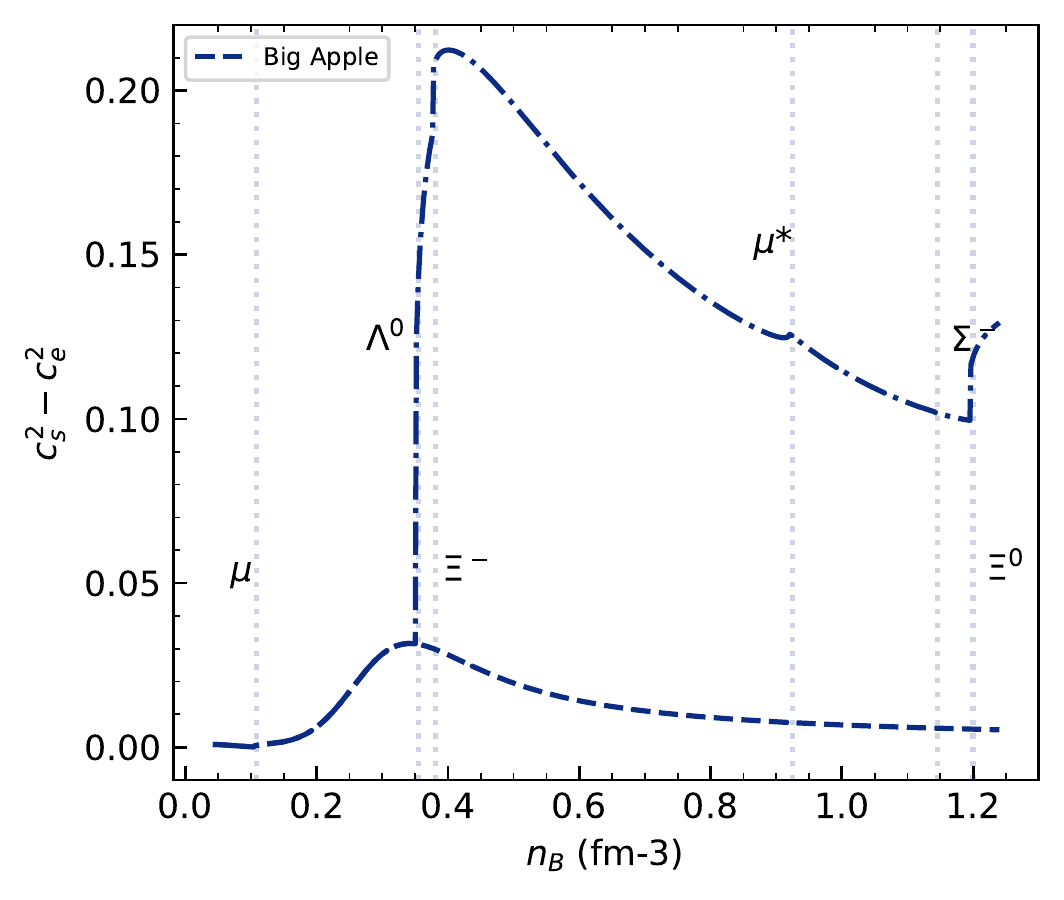}
         \label{fig:big_apple_sound_speed}
     \end{subfigure}\\
     \begin{subfigure}[b]{0.47\textwidth}
         \centering
         \includegraphics[width=\textwidth]{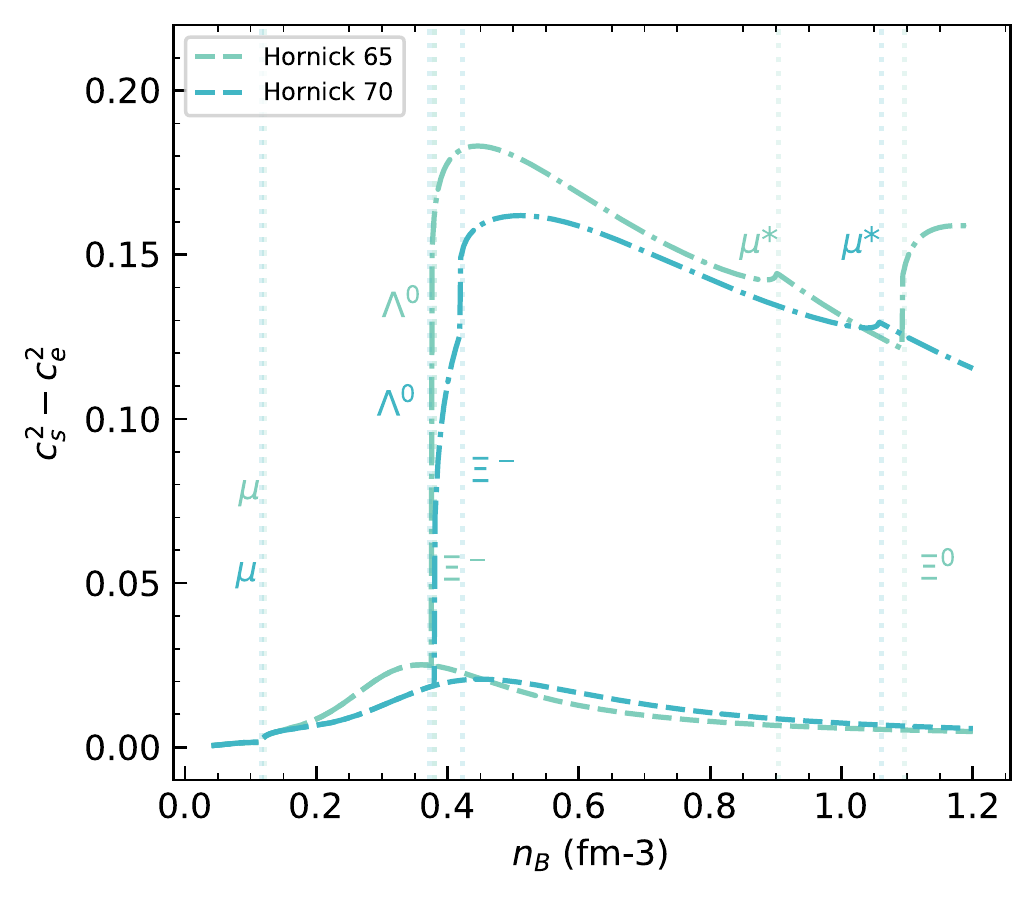}
         \label{fig:hornick_sound_speed_diff}
     \end{subfigure}
     \hfill
     \begin{subfigure}[b]{0.47\textwidth}
         \centering
         \includegraphics[width=\textwidth]{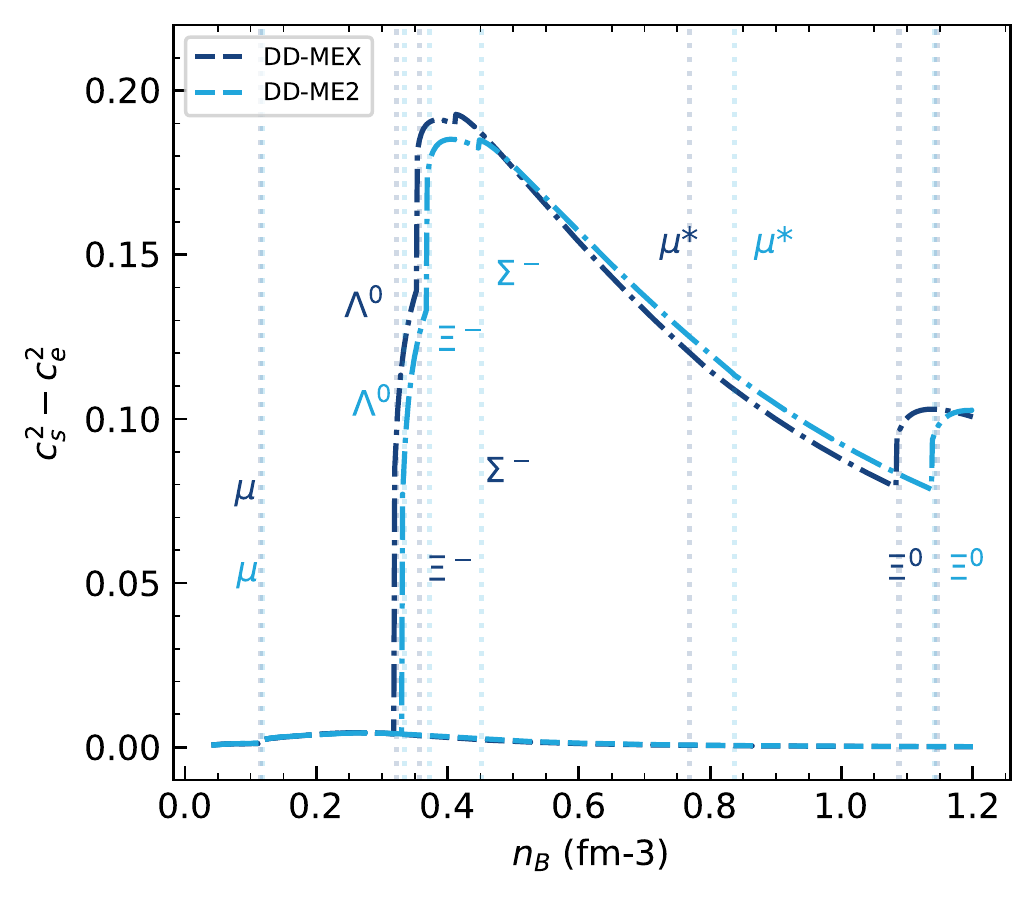}
         \label{fig:dd_rmf_sound_speed_diff}
     \end{subfigure}
    \caption{Sound speed difference $c_s^2 - c_e^2$ plotted for the various models with NPE$\mu$ matter in the dashed lines and NPE$\mu$Y matter in the dot-dashed lines. The vertical dotted lines represent the values of $n_B$ at which new particles emerge corresponding to the ``kinks'' observed in the various curves, except for $\mu^*$ which instead denotes the locations where the muon vanishes. In all models, $c_s^2 - c_e^2$ exhibits a sharp rise upon the emergence of new particles after which $c_s^2 - c_e^2$ begins to decrease until the arrival of a new particle. The appearance of hyperons generally appears to dramatically increase $c_s^2 - c_e^2$. 
    }
    \label{fig:sound_speed_diff}
\end{figure*}

\subsection{Partial Derivative of the Vector Meson Contribution to the Chemical Potential \label{sec:vector_meson}}
The contribution of the vector mesons $\omega, \rho, \phi$ to the chemical potential in the mean-field approximation takes the form
\begin{align}
    \mu_i^{(m)} = g_{\omega i}\omega + I_{3i}g_{\rho i}\rho + g_{\phi i}\phi 
\end{align}
 In similar fashion to sec. \ref{sec:effective_energy}, the partial derivative of $\mu_i^{(m)}$ is dependent on the partial derivatives of the vector meson fields 
\begin{align}
    \frac{\partial \mu_i^{(m)}}{\partial n_B}\bigg|_\chi &= g_{\omega i}\frac{\partial\omega}{\partial n_B}\bigg|_\chi + I_{3i}g_{\rho i}\frac{\partial \rho}{\partial n_B}\bigg|_\chi  + g_{\phi i}\frac{\partial \phi}{\partial n_B}\bigg|_\chi \,.
\end{align}
Each of these partial derivatives of the vector meson fields can be found by differentiating their mean field equations of motion, resulting in 
a system of linear equations for $\partial\omega/\partial n_B$ and $\partial \rho/\partial n_B$ (and $\partial \phi/\partial n_B$) due to the $\Lambda_\omega g_\rho^2 g_\omega^2 \omega^2 \rho^2$ coupling term. In principle, this system of linear equations as written below can be solved exactly, though in our work, we solve them numerically.

\begin{equation} 
\begin{aligned}
    \sum_i g_{\phi i}x_i &= m_\phi^2 \frac{\partial \phi}{\partial n_B} \bigg|_\chi \\
\sum_i g_{\omega i} x_i &= 
    m_\omega^2\frac{\partial \omega}{\partial n_B}\bigg|_\chi + 
    \frac{\xi}{2!}g_{\omega N}^2 \omega^2 \frac{\partial \omega}{\partial n_B} \bigg|_\chi \\
    &\;\; + 2\Lambda_v g_{\rho N}^2 g_{\omega N}^2 \left(2\rho \frac{\partial \rho}{\partial n_B}\bigg|_\chi \omega + \rho^2 \frac{\partial\omega}{\partial n_B}\bigg|_\chi \right)\\
\sum_i g_{\rho i} I_{3i}x_i  &= 
    m_\rho^2 
    \frac{\partial \rho}{\partial n_B} \bigg|_\chi \\
    &\;\; 
    + 2 \Lambda_v g_{\rho N}^2 g_{\omega N}^2 
    \left(\frac{\partial \rho}{\partial n_B} \bigg|_\chi \omega^2 + 2\rho \omega \frac{\partial\omega}{\partial n_B} \bigg|_\chi \right)
\end{aligned}
\end{equation}

% Despite that, we note that we can re-write in matrix form $A\vec{x} = \vec{b}$ where 
% \begin{align}
%     A &= 
%     \begin{pmatrix}
%     m_\omega^2 + 
%     \frac{\xi}{2!}g_{\omega N}^2\omega^2 + \Phi  \rho^2 & 2\Phi \rho \omega \\
%     2 \Phi \rho \omega & m^2_\rho + \Phi  \omega^2 
%     \end{pmatrix}\\
%     x &= 
%     \begin{pmatrix}
%     \frac{\partial \omega}{\partial n_B}\\[6pt]
%     \frac{\partial \rho}{\partial n_B}
%     \end{pmatrix} \qquad 
%     b = 
%     \begin{pmatrix}
%     \sum_i g_{\omega i}x_i\\[6pt]
%     \sum_i g_{\rho i}I_{3i}x_i
%     \end{pmatrix}
% \end{align}
% where $\Phi = 2 \Lambda_v g_{\rho N}^2 g_{\omega N}^2$. 

\begin{figure*}
\begin{center}
    \begin{subfigure}[t!]{0.49\textwidth}
        \includegraphics[width = \textwidth]{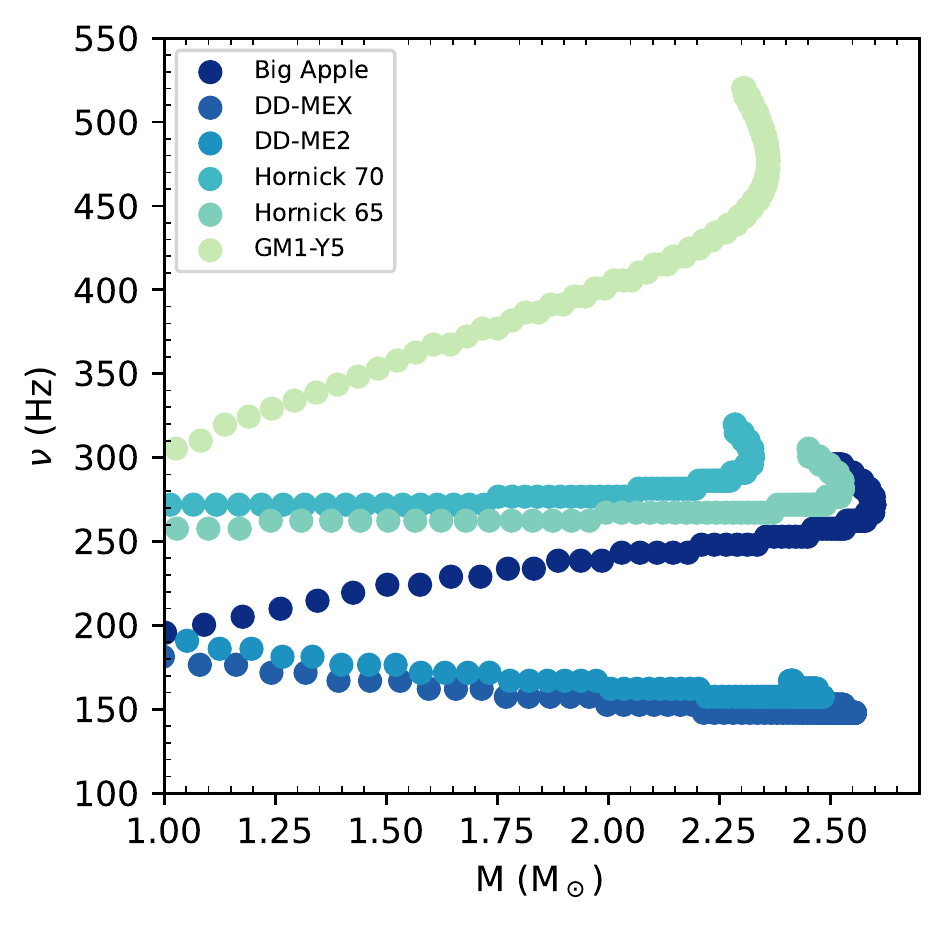}
        \centering 
        \caption{npe$\mu$} 
        \label{fig:g_modes_npemu} 
    \end{subfigure}
    \hfill 
    \begin{subfigure}[t!]{0.49\textwidth}
          \includegraphics[width = \textwidth]{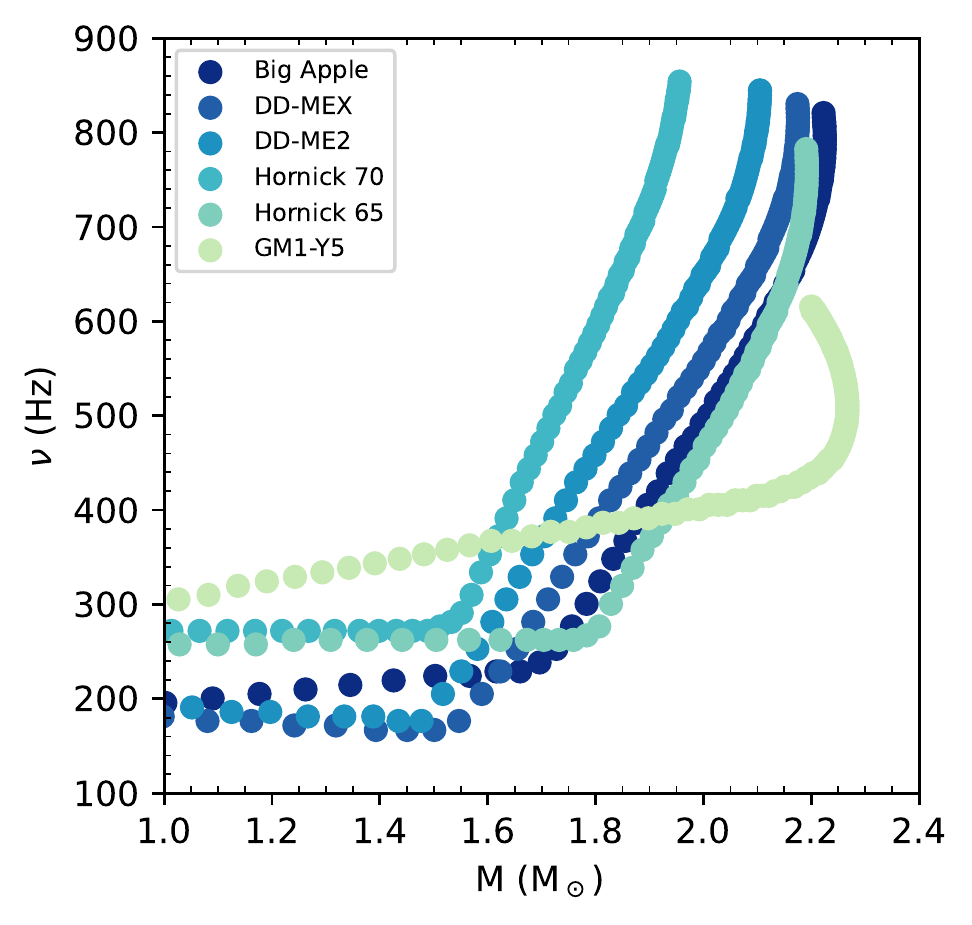}
         \centering 
        \caption{npe$\mu$Y} 
        \label{fig:g_modes_npemuy} 
    \end{subfigure}
    \end{center}
    \caption{$g$-mode oscillation frequency as a function of stellar mass for npe$\mu$ composition on the left and npe$\mu$Y composition on the right. As a result of dependence of the sound speed difference on the number of equilibrating species in the system, the \gm\,frequency rises sharply when the threshold density for a new species that participates in $\beta$-equilibrium reactions is breached. The case of GM1-Y5 is markedly different from the other models: the difference arises due to the absence of quartic interactions or SU(6) coupling constants, which forces hyperons to appear only at the tail end of the mass-radius curve.}
    \label{fig:g_modes}
\end{figure*} 

\newpage 

\subsection{DDRMF Model Modifications}

For the DDRMF models, as the coupling constants are density dependent, the contribution to the partial derivative of the effective mass are given by 
\begin{align}
    \frac{\partial m_i^*}{\partial n_B}\bigg|_\chi  &= - \frac{\partial g_{\sigma i}}{\partial n_B}\bigg|_\chi  \sigma - g_{\sigma i}
    \frac{\partial \sigma}{\partial n_B}\bigg|_\chi -
    \frac{\partial g_{\xi i}}{\partial n_B}\bigg|_\chi \xi\\
    &\qquad - 
    g_{\xi i}\frac{\partial \xi}{\partial n_B}\bigg|_\chi - \frac{\partial g_{\delta i}}{\partial n_B}\bigg|_\chi \delta 
    - g_{\delta i}\frac{\partial \delta}{\partial n_B}\bigg|_\chi 
\end{align}
and the partial derivative of the mesonic contribution to the baryon chemical potential are given by 
\begin{align}
    \frac{\partial \mu_i^{(m)}}{\partial n_B}\bigg|_\chi &= \frac{\partial g_{\omega i}}{\partial n_B}\bigg|_\chi \omega + 
    g_{\omega i} \frac{\partial \omega}{\partial n_B}\bigg|_\chi + I_{3i}\frac{\partial g_{\rho i}}{\partial n_B}\bigg|_\chi\rho\\
    &\qquad + I_{3i}g_{\rho i}\frac{\partial \rho}{\partial n_B}\bigg|_\chi + \frac{\partial g_{\phi i}}{\partial n_B}\bigg|_\chi\phi + 
    g_{\phi i}\frac{\partial \phi}{\partial n_B}\bigg|_\chi
\end{align}
where as usual, the partial derivatives with respect to $n_B$ are taken at fixed composition. The mesonic equations of motion are likewise modified, though the overall structure of the resulting equations, and hence the solution methods, are no more complicated than for the NLRMF models.

The additional re-arrangement term $\Sigma^r$ can be differentiated in a similar manner. However, we note that in our context, we are ultimately interested in $\tilde{\mu_i}$ which by its dependence on the difference of the neutron and baryon chemical potentials (as in eqn. (\ref{mu_tilde}))  leads to the contributions from $\partial \Sigma^r/\partial n_B$ from the neutron and $i$th baryon cancelling each other out and ultimately does not contribute to $c_s^2 - c_e^2$.

% \begin{figure}[H]
%     
%     \includegraphics[scale=0.85]{Sound_Speed_Diff/all_sound_diff_gm1_new_scaled.pdf}
%     \caption{Caption}
%     \label{fig:gm1_y5_sound_speed}
% \end{figure}

% \begin{figure}[H]
%     \centering
%     \includegraphics[scale=0.85]{Sound_Speed_Diff/RMF_GLUCK.pdf}
%     \caption{Caption}
%     \label{fig:rmf_gluck}
% \end{figure}

% \begin{figure}[H]
%     \centering
%     \includegraphics[scale=0.85]{Sound_Speed_Diff/big_apple_npe_soundspeeddiff_sq.pdf}
%     \caption{Sound Speed Difference for the Big Apple model where the solid, dashed, and dash-dotted lines represent NPE, NPE$\mu$, and NPE$\mu$Y matter respectively. The vertical lines indicate at which values of $n_B$ a new particle populates. This corresponds to the ``kinks'' in the $c_s^2 - c_e^2$ curve. In contrast, $\mu^*$ indicates the location at which the $\mu$ vanishes. We see that with the onset of $\Lambda^0$ that $c_s^2 - c_e^2$ spikes up dramatically}
%     \label{fig:sound_speed_difference_big_apple}
% \end{figure}

% \begin{figure}[H]
%     \centering
%     \includegraphics[scale=0.85]{Sound_Speed_Diff/DD-ME2_sound_speed.pdf}
%     \caption{Caption}
%     \label{fig:my_label}
% \end{figure}

% \subsection{Sound Speed Difference}
% \textcolor{red}{Is this not basically Appendix A?}

\newpage

\section{Results \label{sec:results}}

The sound speed difference for the models considered in this work are collected in the panels of Fig. \ref{fig:sound_speed_diff}. A common observation is that the sound speed difference experiences a sharp rise when a new species threshold is breached, due to a drop in $c_e^2$. This effect is quite dramatic for hyperons, particularly the $\Lambda^0$. The gradual decrease of the sound speed difference between consecutive species thresholds signifies that the system is returned to chemical and mechanical equilibrium. A comparison to $npe\mu$ matter alone highlights the remarkable effect of hyperons on the sound speed difference. Muons, due to their relatively small fraction compared to hyperons (see Fig. \ref{fig:big_apple_frac}), do not impact the sound speed as much as hyperons. From the hyperon species, the $\Lambda$ has the largest relative effect due to its population fraction.

There are more subtle differences, as reflected in $\mu^{\ast}$, between the various models as well, due to variations in the baryon-meson, meson-meson interactions in the Lagrangian, the nature of the coupling constants (density-dependent or not), as well as the recipe chosen to fix meson-hyperon couplings. 

% \textcolor{red}{First, the DDRMF models predict a comparatively much smaller peak $npe\mu$ sound speed difference when compared to..}

% \textcolor{red}{In particular, the GM1-Y5 model..}
% \textcolor{blue}{Unless you know the reason for a particular feature in the graph, no need to point it out. If referee asks, we can delve deeper.}

The implication of these trends in the sound speed difference is that the $g$-mode frequency, through the \bv frequency, would be expected to manifest similar dramatic features for $npe\mu Y$ compositions. Indeed, our results for the \gm oscillations presented in Fig. \ref{fig:g_modes} demonstrate this fact. Specifically, a comparison of Figs. \ref{fig:g_modes_npemu} and \ref{fig:g_modes_npemuy}  for $npe\mu$ compositions and $npe\mu Y$ respectively (for each of the six RMF models used in this work), show that in all but one of the $npe\mu Y$ models (GMI-Y5), a dramatically sharp increase in the oscillation frequency occurs at around 1.5-1.6$\,M_\odot$. This corresponds to the lightest hyperon threshold in the star. The \gm\,frequencies for the stars with $npe\mu Y$ composition are approximately 350-\SI{750} Hz larger, depending on the stellar mass, than for those with $npe\mu$ composition. The case of GM1-Y5 is markedly different due to the absence of quartic interactions or SU(6) coupling constants, pushing the threshold density of hyperons near the tail end of the mass-radius curve. \\

This qualitative behavior of the $g$-mode frequency upon the onset of new degrees of freedom is similar to results in \cite{PJ-PRD,Zhao:2022toc,Constantinou_2021} where a transition to quark matter in the form of a mixed/crossover quark matter phase was considered. In that case, the principal core \gm\, frequency for hybrid stars containing quark matter was in the range  $\approx$ 200 - 600 Hz,  and therefore less dramatic than the the effect of hyperons. Since the frequencies of stars without strangeness degrees of freedom is only about 100 - 200 Hz, we conclude that a precise determination of the \gm\, frequency if and when observed in perturbed neutron stars, could potentially be a signature of strangeness, but also allow us to discern if such strangness is bound (hyperons) or free (quarks).

% \begin{figure}
%     \centering
%     \includegraphics[scale=0.8]{gmode_oscillations/g_mode_gm1_y5.pdf}
%     \caption{Caption}
%     \label{fig:gm1_y5_gmode}
% \end{figure}

% \begin{figure}
%     \centering
%     \includegraphics[scale=0.8]{gmode_oscillations/g_mode_rmf_gluck.pdf}
%     \caption{Caption}
%     \label{fig:rmf_gluck}
% \end{figure}

% \begin{figure}
%     \centering
%     \includegraphics[scale=0.8]{gmode_oscillations/g_mode_big_apple.pdf}
%     \caption{Caption}
%     \label{fig:g_mode_big_apple}
% \end{figure}

% \begin{figure}
%     \centering
%     \includegraphics[scale=0.8]{gmode_oscillations/g_mode_dd_mex.pdf}
%     \caption{Caption}
%     \label{fig:g_mode_dd_mex}
% \end{figure}

% \begin{figure}
%     \centering
%     \includegraphics[scale=0.8]{gmode_oscillations/dd_me2.pdf}
%     \caption{Caption}
%     \label{fig:g_mode_dd_me2}
% \end{figure}

\section{Conclusions \label{sec:conclusions}}

The main objective of this work was to ascertain the characteristics of $g$-mode oscillations of hyperonic stars, comparing them to the standard $npe\mu$  composition of a neutron star. Toward this end, we used a variety of relativistic mean field approaches to model the core of the star, where hyperons can be present. In particular, we used models GM1-Y5 \cite{Oertel_2015, Glendenning_1991}, Big Apple \cite{Das_2021, Fattoyev_2020}, Hornick 65, 70 \cite{Hornick_2018}, DD-MEX \cite{Taninah_2020, Tu_2022, Thapa_2021, Huang_2020}, and DD-ME2 \cite{lalazissis_2005, Tu_2022, Thapa_2021, Huang_2020}. The models were chosen to sample a variety of different possible baryon-meson and meson-meson interactions as well as include different treatments of the coupling constants, including models where the coupling constants vary with total baryon number density (DDRMF). All models satisfy current astrophysical constraints, producing equations of state stiff enough to produce maximally sized stars as well as constraints on the mass-radius relations in agreement with NICER constraints on PSR J0030+0451 and PSR J0740+6620. The calculated tidal deformabilities also agree with current constraints placed by GW170817. 

While $M$-$R$ curves only depend on the pressure vs density relation (EOS),  the analysis of \gm~oscillations requires simultaneous information about the equilibrium and adiabatic squared sound speeds, $c_e^2=dp/d\varepsilon$ and 
$c_s^2=\partial p/\partial\varepsilon|_x$, where $x$ are the local, independent composition variables.  The distinction between these two sound speeds plays a central role in determining the \bv~frequencies $\omega^{2} \propto c_e^{-2} - c_s^{-2}$ of non-radial \gm~oscillations. We generalized the method applied in~\cite{1994MNRAS.270..611L,PJ-PRD} for $npe\mu$ matter to calculate the sound speed difference $c_s^2 - c_e^2$ from partial derivatives of linear combinations of chemical potentials, and applied this to obtain the $g$-mode spectrum for hyperonic stars described by relativistic mean field models. 

We find that the $g$-mode is sensitive to the presence of hyperons in neutron stars, as signalled by the sharp changes in sound speed difference at the lightest hyperon threshold (Fig. 6), raising the local \bv\, frequency and the fundamental \gm\, frequency of the star (Fig. 7). Contrasts of \gm\ frequencies between normal and hyperonic stars containing quark matter (Fig. 7) form the principal results of our work. This constrast is a common feature that arises across the different models of hyperpnic matter, and gives confidence that the effect is representative of the change in composition rather than an artifact of a specific model. 

The novel feature of this work is the first calculation of the two sound speeds in hyperonic matter and its impact on the principal $g$-mode frequency of hyperonic stars. Our results suggests that determining the composition of the star through $g$-modes is a possible resolution to breaking degeneracies in inferences on the equation of state from $M$-$R$ data alone, and ascertain if strangeness exists in neutron stars. Future work is aimed at quantifying the $g$-mode frequencies for hyperonic stars with a phase transition to quark matter or crossover transitions as in quarkyonic matter. It would also be interesting to study the evolution of the $g$-mode in binary mergers where one or both components may be a hyperonic star, since such modes can be excited during inspiral and potentially alter the phase and amplitude of the gravitational wave signal from coalescing ordinary neutron stars.

%%%%%%%%%%%%%%%%%%%%%%%%%%

 \section*{Acknowledgements}
V.T. and P.J. are supported by the U.S. National Science Foundation Grant PHY-1913693. 
\newpage

\appendix

\section{Demonstrating Validity of Sound Speed Difference Expression \label{sound_speed_diff_proof}}

It was shown in \cite{PJ-PRD} that from the definitions of $c_s^2$ and $c_e^2$ that the sound speed difference $c_s^2 - c_e^2$ could be re-written as 
\begin{align}
    c_s^2 - c_e^2 = \frac{1}{\mu_\text{avg}}\frac{\partial p}{\partial n_B}\bigg|_\chi - 
    \frac{1}{\mu_n}\frac{dp}{dn_B} \label{cs2_ce2_orig}
\end{align}
where $\mu_\text{avg} := \sum_i \mu_i x_i$, $\mu_n$ is the neutron chemical potential and $\partial p/\partial n_B|_\chi$ is the partial derivative of pressure with respect to baryon density $n_B$ while holding composition fixed. 
This expression was then shown to be able to be re-written in terms of partial derivatives of $\tilde{\mu}_i$ for the specific case of $npe$ and $npe\mu$ matter where the independent variables chosen were the electron fraction $x_e$ in the first case and the lepton fraction $x$ and muon fraction $y$ in the second case. Then the sound speed difference in the $npe\mu$ case was able to be re-written as 
\begin{align}
    c_s^2 - c_e^2 = 
    - \frac{n_B^2}{\mu_n}
    \left(\frac{\partial\tilde{\mu}_x}{\partial n_B}\bigg|_{x,y}\frac{dx}{dn_B} + 
    \frac{\partial \tilde{\mu}_y}{\partial n_B}\bigg|_{x,y} \frac{dy}{dn_B}\right)
\end{align}

Here, following the same steps outlined in \cite{PJ-PRD}, we can generalize these results to any arbitrary composition of baryons and leptons to get the expression shown in equation \ref{cs2_ce2} starting from equation \ref{cs2_ce2_orig}.   

We can start by taking the neutron fraction $x_n$ and the electron fraction $x_e$ to be the dependent fractions for all compositions. This then implies that all other baryon and lepton fractions are independent variables in our system. This type of scheme has the advantage of allowing us to write a generalized expression for sound speed difference for a variety of different compositions that may occur as $n_B$ increases and heavier particles such as hyperons appear without having to re-define and re-solve for different independent and dependent fractions. 

Then the pressure $p = p(n_B, x_1, \hdots, x_n)$ is a function of total baryon density $n_B$ and the independent baryon, lepton fractions $x_1,\hdots, x_n$ so the total derivative of $p$ with respect to $n_B$ is given by
\begin{align}
    \frac{dp}{dn_B} = 
    \frac{\partial p}{\partial n_B}\bigg|_\chi 
    + \sum_i \frac{\partial p}{\partial x_i}\bigg|_{n_B, x_j \neq x_i} \frac{dx_i}{dn_B}
\end{align}
where the sum over $i$ is over all independent baryon and lepton fractions/particles. When inserted into equation \ref{cs2_ce2_orig} we can expand and collect terms in the following manner
\begin{align}
    c_s^2 - c_e^2 &= 
    \left(\frac{1}{\mu_\text{avg}} - \frac{1}{\mu_n}\right)
    \frac{\partial p}{\partial n_B}\bigg|_{n_B, x_j \neq x_i}\\
    &\quad - 
    \frac{1}{\mu_n} \sum_i 
    \left(\frac{\partial p}{\partial x_i}\bigg|_{n_B, x_j \neq x_i}\frac{dx_i}{dn_B}\right)\\
    &= \left(\frac{\mu_n - \mu_\text{avg}}{\mu_\text{avg} \cdot \mu_n}\right)\frac{\partial p}{\partial n_B}\bigg|_\chi\\
    &\qquad - 
    \frac{1}{\mu_n}\sum_i\left(\frac{\partial p}{\partial x_i}\bigg|_{n_B, x_j \neq x_i}\frac{dx_i}{dn_B}\right)
\end{align}

Next, the average chemical potential $\mu_\text{avg}$ can be expanded as 
\begin{align}
    \mu_\text{avg} &:= \sum_j \mu_j x_j\\
    &= x_n \mu_n + x_e \mu_e + \sum_i x_i \mu_i \qquad i \in \text{ind. var}
\end{align}
But with the neutron and electron fractions as dependent variables, we can re-write them in terms of the other independent fractions using the constraints of charge neutrality and baryon number conservation. 

\begin{align}
    1 &= x_n + \sum_b x_b \qquad b \in \text{baryon}\\
    0 &= -x_e - x_\mu + \sum_b q_b x_b 
\end{align}
After solving for $x_n$ and $x_e$ in terms of the other fractions using these two constraints $\mu_\text{avg}$ becomes 
\begin{align}
    \mu_\text{avg} &= \left(1 - \sum_b x_b\right)\mu_n + \left(-x_\mu + \sum_b q_b x_b\right)\mu_e\\
    &\qquad + x_\mu \mu_\mu + \sum_b x_b \mu_b
\end{align}

Then the difference $\mu_n - \mu_\text{avg}$ becomes 
\begin{align}
    \mu_n - \mu_\text{avg} &= 
    \sum_b x_b \mu_n - \sum_b x_b\mu_b - \sum_b q_b x_b \mu_e\\
    &\qquad + x_\mu (\mu_e - \mu_\mu)\\
    &= \sum_b (\mu_n - q_b \mu_e - \mu_b) x_b\\
    &\qquad + x_\mu (\mu_e - \mu_\mu)
\end{align}
But we see that the terms inside of the parentheses are exactly combinations of chemical potentials that vanish in $\beta$ equilibrium
\begin{align}
    \tilde{\mu}_b &= \mu_n - q_b \mu_e - \mu_b = 0\\
    \tilde{\mu}_\mu &= \mu_e - \mu_\mu 
\end{align}
which allows us to re-write $\mu_\text{avg}$ in a concise manner in terms of $\tilde{\mu}_i$
\begin{align}
    \mu_n - \mu_\text{avg} &= 
    \sum_b \tilde{\mu}_b x_b + \tilde{\mu}_\mu x_\mu\\
    &= \sum_i \tilde{\mu}_i x_i \qquad i \in \text{all ind. vars}
\end{align}

The remaining steps follow in a similar fashion as described in \cite{PJ-PRD}. In $\beta$ equilibrium $\mu_n - \mu_\text{avg}$ is zero since $\tilde{\mu}_i = 0$ for all $i$. The sound speed difference expression reduces to 
\begin{align}
    c_s^2 - c_e^2 =
    - \frac{1}{\mu_n}
    \sum_i \frac{\partial p}{\partial x_i}\bigg|_{n_B,x_j\neq x_i}
    \frac{dx_i}{dn_B}
\end{align}
Using $p = n_B^2 \partial E/\partial n_B|_{\chi}$ we can re-write this as
\begin{align}
     c_s^2 - c_e^2 &=
    - \frac{n_B^2}{\mu_n}
    \sum_i \frac{\partial E}{\partial x_i}\bigg|_{n_B,x_j\neq x_i}
    \frac{dx_i}{dn_B}\\
    &= 
    \frac{n_B^2}{\mu_n}
    \sum_i \frac{\partial \tilde{\mu}_i}{\partial n_B}\bigg|_\chi \frac{dx_i}{dn_B} \label{eqn:sound_speed_diff_appendix}
\end{align}
where eqn. (\ref{eqn:sound_speed_diff_appendix}) is the expression we use in calculating the sound speed difference in this paper.

\section{Mesonic Mean Field Equations \label{sec:mesonic_mean_field}}
The form of the Euler-Lagrange field equations for the mesons as specified for the general Lagrangian used in our work, that is, including the form of the meson-meson interactions as given in eqn. (\ref{eqn:U_NL}).

\begin{align}
    m_\sigma^2 \sigma + 
  b g_{\sigma N}^3 \sigma^2 + 
  c g_{\sigma N}^4 \sigma^3
    &= \sum_i g_{\sigma i}n_i^s \label{eqn:sigma_eom}\\
    m_\omega^2 \omega 
    + 
    \frac{\xi}{3!}g_{\omega N}^2 \omega^3 
    + 2 \Lambda_\omega g_{\rho N}^2 g_{\omega N}^2 \rho^2 \omega &= \sum_i g_{\omega i} n_i \label{eqn:omega_eom}\\
        m_\rho^2 \rho 
    + 2 \Lambda_\omega g_{\rho N}^2 g_{\omega N}^2 \rho \omega^2 &= \sum_i g_{\rho i} I_{3i} n_i \label{eqn:rho_eom}\\
    m_\phi^2\phi &= \sum_i g_{\phi i}n_i \label{eqn:phi_eom}\\
    m_\xi^2 \xi &= \sum_i g_{\xi i}n_i^s \label{eqn:xi_eom}\\
    m_\delta^2 \delta &= \sum_i I_{3i} g_{\delta i}n_i^s \label{eqn:delta_eom}
\end{align}

%\section{Particle Fractions}

\nocite{*}

\bibliography{bib}% Produces the bibliography via BibTeX.

\end{document}